\documentclass[aps,pre,twocolumn,amsfonts,amssymb,amsmath,showpacs,floatfix]{revtex4}

\usepackage{graphics}
\usepackage{dcolumn}

\newcommand{\conjugate}[1]{\bar{#1}}
\renewcommand{\exp}[1]{e^{#1}}
\renewcommand{\vec}[1]{\mathbf{#1}}
\newcommand{\mat}[1]{\mathbf{#1}}
\newcommand{\order}{\mathcal{O}}
\newcommand{\degree}{^{\circ}}
\newcommand{\re}{\mathop{\mathrm{Re}}}

\newcommand{\Dhat}{\mathop{\widehat{\mathcal{D}}}}
\newcommand{\sgn}{\mathop{\mathrm{sign}}}

\newcommand{\ratio}[2]{#1\colon\negthinspace#2}

\begin{document}
\title{Multi-Frequency Control of {F}araday Wave Patterns}

\author{Chad M. Topaz}
\affiliation{Department of Mathematics, UCLA, Los Angeles,  
California 90095, USA}
\email{topaz@ucla.edu}
\author{Jeff Porter}
\affiliation{Department of Applied Mathematics, University of Leeds,  
Leeds LS2 9JT, UK}
\author{Mary Silber}
\affiliation{Department of Engineering Sciences and Applied  
Mathematics, Northwestern University, Evanston, Illinois 60208, USA}

\date{\today}

\begin{abstract}

We show how pattern formation in Faraday waves may be manipulated 
by varying the harmonic content of the periodic forcing function.  Our approach 
relies on the crucial influence of resonant triad interactions coupling 
pairs of critical standing wave modes with damped, spatio-temporally resonant 
modes.  Under the assumption of weak damping and forcing, we perform a 
symmetry-based analysis that reveals the damped modes most relevant for 
pattern selection, and how the strength of the corresponding triad interactions 
depends on the forcing frequencies, amplitudes, and phases. In many cases, the  
further assumption of Hamiltonian structure in the inviscid limit determines 
whether the given triad interaction has an enhancing or suppressing effect 
on related patterns.  Surprisingly, even for forcing functions with arbitrarily 
many frequency components, there are at most five frequencies that affect each of the important triad interactions at leading order. The relative 
phases of those forcing components play a key role, sometimes making the 
difference between an enhancing and suppressing effect.  In numerical examples, 
we examine the validity of our results for larger values of the  damping and forcing.  
Finally, we apply our findings to one-dimensional periodic patterns obtained with impulsive forcing and to two-dimensional superlattice patterns and quasipatterns obtained with multi-frequency forcing.
\end{abstract}

\pacs{05.45.-a, 47.35.+i, 47.54.+r, 89.75.Kd}

\maketitle

\section{Introduction \label{sec:introduction}}

Parametrically forced surface waves have proven to be a rich and versatile 
source of patterns since their initial observation by Michael Faraday in 
1831 \cite{f1831}. These Faraday wave  patterns are composed of 
standing waves set up in response to periodic vertical vibration of 
sufficient strength. Early investigations (see \cite{mh1990,mfp1998} for 
reviews) used a sinusoidal forcing function and focused on simple patterns 
such as stripes, squares, and hexagons, which oscillate in subharmonic 
response to the forcing.   Recently, experimentalists have used multi-frequency 
forcing to generate more complex states such as quasipatterns and superlattice 
patterns \cite{m1993,ef1993,ef1994,kpg1998,af1998,af2000a,af2000b,af2002}.
These observations have, in turn, fueled theoretical interest in such patterns 
and in multi-frequency forcing 
\cite{bet1996,zv1997b,lp1997,sp1998,sts2000,trhs2000,ts2002,ps2002}. 

The use of multi-frequency forcing requires the selection of a large number 
of control parameters.  The forcing frequencies, their amplitudes, and their 
relative phases may all affect the pattern formation problem in a nontrivial 
way \cite{m1993,zv1997b,ps2002,pts2004}.  Further complexity arises 
from the presence of multiple length scales.  In addition to the length scales 
driven by the various forcing frequencies in accordance with the
 parametric (\emph{i.e.}, subharmonic) resonance conditions, there are 
 numerous damped modes that are driven {\it nonlinearly}.  Many of these 
 can significantly influence the dynamics of the critical modes.  For instance, 
 in the case of two-frequency forcing, the damped mode that oscillates at 
 the so-called difference frequency is important for selecting superlattice 
 patterns~\cite{ts2002}.

Resonant triad interactions -- the lowest order nonlinear interactions -- 
provide a useful framework for investigating the relationship between 
the many control parameters and length scales in the multi-frequency 
forced Faraday wave problem.  Resonant triads that couple two critical 
modes with a damped, spatiotemporally resonant mode play a key role in 
the nonlinear pattern selection process.  Most of these damped modes 
function as energy sinks, effectively creating an anti-selection mechanism 
that suppresses the triad interaction and thereby favors patterns which avoid 
the corresponding resonant angle.  However, other damped modes 
act as energy sources, providing a positive selection mechanism 
that helps stabilize patterns involving the associated resonant angle.  
The effect of different damped modes on pattern selection is investigated 
in \cite{pts2004}, which, for forcing functions with up to three frequency 
components, determines the most important damped modes, their 
effect (enhancing or suppressing) on associated patterns, and the 
dependence of the nonlinear interaction on the forcing frequencies, 
amplitudes, and relative phases.   These results are used to interpret 
recent Faraday wave experiments that produced complex patterns, 
namely, a two-frequency forced superlattice pattern in \cite{kpg1998} 
and a three-frequency forced quasipattern in \cite{af2002}.
The approach developed in \cite{pts2004} follows from a systematic 
consideration of weakly broken symmetries: time translation, time reversal,
and Hamiltonian structure (see \cite{ps2002}), and is therefore most 
relevant for systems with weak damping and forcing. In this limit 
the vastness of the control parameter space can be an asset, enabling 
one to enhance or suppress particular triad interactions simply by 
tuning the appropriate forcing parameters.   

In this paper, we adopt the same prescriptive approach to Faraday wave
pattern formation, describing in more detail the technique for exploiting 
weakly broken symmetries, and extending the results of \cite{pts2004} to 
forcing functions containing \emph{arbitrarily many} Fourier components. 
We determine which damped modes are favored by a strong nonlinear coupling 
and tabulate how the corresponding resonant triad interactions depend on the 
forcing parameters.  A somewhat surprising result, which makes this project 
feasible, is that for a given damped mode there are at most five out of the 
potentially infinite number of forcing frequency components in the forcing function
that affect the resonant triad interaction at leading order in the damping 
parameter~$\gamma$ (defined below).    We investigate numerically the validity 
of our predictions with respect to the small $\gamma$ assumption.   This is important 
for understanding the extent to which the symmetry-based picture 
we develop here can be applied to realistic experiments.   We then use several 
different numerical examples to illustrate how the resonant triad interactions most 
relevant to pattern formation may be controlled through a judicious choice 
of forcing parameters.

The remainder of this paper is organized as follows. In Section~\ref{sec:background}, 
we review basic ideas concerning the importance of resonant triad interactions to 
Faraday wave pattern formation, including a discussion of some of the previous 
theoretical and experimental work.  Section~\ref{sec:symmetry} contains our 
symmetry-based analysis.  We enumerate the most important weakly damped modes, 
calculate their effect on pattern formation, and determine the dependence of this effect 
on the forcing parameters. Section~\ref{sec:discussion} contains a general discussion of the symmetry-based 
results.   We study their range of validity with respect to $\gamma$ by comparing 
the symmetry-based predictions to numerical results obtained using the 
Zhang-Vi\~{n}als Faraday wave equations \cite{zv1997a}.  In Section~\ref{sec:applications}, 
we apply our symmetry-based results in several examples. In the first 
application, we consider weakly-nonlinear periodic patterns forced by a repeated 
sequence of $\delta$-functions of alternating sign.  In accordance with the results first reported in~\cite{cps2004}, we demonstrate how, by 
varying the spacing between the pulses, we may control the amplitude of the 
pattern. In the second application, we show how to construct a five-frequency 
forcing function which leads to dramatic stabilization of a complex pattern, namely 
an SL-I superlattice pattern of the type observed in~\cite{kpg1998}.  In the third 
example, we conjecture about a seven-frequency forcing function which should  
be conducive to the experimental observation (as yet, lacking) of 14-fold 
quasipatterns.  We summarize and conclude in Section~\ref{sec:conclusions}.

\section{Background \label{sec:background}}

We lay the groundwork for our new results by reprising basic ideas from 
\cite{ss1999,sts2000,ts2002,ps2002,ps2003,pts2004} on the role of resonant 
triads in Faraday wave pattern formation. We consider Faraday waves on an 
unbounded horizontal domain subjected to an arbitrary periodic forcing 
function $f(t)$.  We use a dimensionless time $\tau$ such that the common 
frequency is one, and expand $f(\tau)$ in a Fourier series: 
\begin{equation}
\label{eq:f(t)}
f(\tau) = \sum_{u \in \mathbb{Z}^{+} } f_u \exp{i u \tau}+c.c.,
\quad f_u \in \mathbb{C}
\end{equation}
where $u = m,n,p,\ldots$ are the forcing frequencies (distinct and co-prime), 
$|f_u|$ are the forcing amplitudes, and $\phi_u = \arg(f_u)$ are the 
corresponding phases.   Without loss of generality, we take $m$ to be the  
``dominant'' frequency, \emph{i.e.} we assume that $f_m$ (to lowest order) is 
the component that drives the critical modes (this does not necessarily mean 
that $|f_m|$ is the largest of the $|f_u|$). There exists a bifurcation point 
$|f_m| = |f_m|^{crit}$ which depends on the physical properties of the
fluid, and on the other $f_u$, below which the flat fluid state is stable to 
perturbations of all wave numbers, and at which perturbations of (generically) 
one critical wave number $k_c$ become neutrally stable.  We consider the 
properties of resonant triads in a vicinity of this bifurcation in 
parameter space.

\begin{figure}[ht]
\centerline{\includegraphics{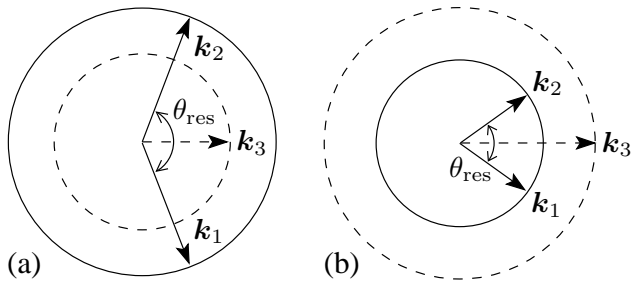}} 
\caption{Fourier space diagram of spatially resonant triads satisfying 
(\ref{eq:resonance-condition}).  The two neutrally stable modes have wave number
$|\vec{k}_1|=|\vec{k}_2|=k_c$ and oscillate with dominant frequency $m/2$. 
The damped mode has $|\vec{k}_3| = k_d$ and oscillates with dominant 
frequency $\Omega$. (a) $k_d < k_c$. (b) $k_c < k_d < 2k_c$.}  
\label{fig:resonant_triad}
\end{figure}

Three wave, or triad, resonance is the simplest nonlinear mechanism by 
which different waves may interact.   The three waves involved have Fourier 
wave vectors $\vec{k}_j$, $j=1,2,3$, satisfying
\begin{equation}
\vec{k}_1 + \vec{k}_2 = \vec{k}_3. \label{eq:resonance-condition}
\end{equation}
In this paper we are interested in the influence of the damped modes 
that are driven nonlinearly (through resonant triad interaction) by the 
critical modes.  Hence two of the wave vectors have the critical value 
$|\vec{k}_1|=|\vec{k}_2|=k_c$.  These waves, to first approximation, 
respond subharmonically to the dominant forcing component $m$ and thus 
oscillate with predominant frequency $m/2$.  The third wave in the triad 
has wave number $|\vec{k}_3| = k_d$ and is associated with a damped mode 
with dominant frequency $\Omega$.   The values of $\Omega$ most relevant 
to Faraday wave pattern formation are determined in Section~\ref{sec:symmetry}.   
The condition~(\ref{eq:resonance-condition}) defines an angle of spatial 
resonance $\theta_{\rm res} \in [0,\pi)$ between the two critical modes:
\begin{equation}
\cos \frac{\theta_{\rm res}}{2}  = \frac{k_d}{2 k_c}.
\label{eq:resonant-angle}
\end{equation}
We exclude the cases $\theta_{\rm res}=\pi/3$ and $\theta_{\rm res}=2\pi/3$ 
since these correspond to hexagons and $k_d$ would then not be 
damped.   Figure~\ref{fig:resonant_triad} shows Fourier space diagrams 
corresponding to the resonant triad described above.
%%%%%%%%%%%%%%%%%%%%%%%
%%%%%%%%%%%%%%%%%%%%%%%

In the presence of damping, the primary Faraday instability leads to 
standing waves (SW).  We associate with the wave vectors $\vec{k}_j$ 
three complex amplitudes $A_j$ which describe the slow-time evolution 
of the three standing wave modes pictured in Figure~\ref{fig:resonant_triad} 
(the fast-time subharmonic oscillation of the waves has been factored out; see, 
\emph{e.g.}, \cite{sts2000}). Evolution equations for the $A_j$ can be 
obtained by applying a standard reduction procedure to the governing 
equations, as we shall do in Section~\ref{sec:applications}.  However, 
at this point we are concerned only with the form of these equations, 
which is determined by the spatial symmetries.  The action of spatial 
translation is
\begin{gather}
\label{eq:SWspatialtrans}
T_{\Theta}: A_j \rightarrow A_j \exp{i\theta_j}, \\*
\Theta = (\theta_1,\theta_2),\quad \theta_1,\theta_2 \in [0,2\pi),  
\quad \theta_3 = \theta_1+\theta_2, \nonumber 
\end{gather}
%]
while a reflection about $\vec{k}_3$ leads to  
\begin{equation}
\mathcal{\kappa}: A_1 \leftrightarrow A_2,  \label{eq:SWspatialreflec2}
\end{equation}
and a rotation by $\pi$ acts as
\begin{equation}
\mathcal{R}: A_j \rightarrow \conjugate{A}_j.  
\label{eq:SWspatialreflec1}
\end{equation}
Equivariance under these three symmetries (see, \emph{e.g.}, \cite{gss1988}) 
requires that the differential equations describing the dynamics of the $A_j$ 
take the form 
\begin{subequations}
\begin{eqnarray}
\label{eq:standing-wave-eqns}
\dot{A_1} & = & \Lambda_1 A_1 + \alpha_1 \overline{A}_2 A_3 \\
& & \mbox{}+ (a |A_1|^2+b_0 |A_2|^2+b_1|A_3|^2)A_1, \nonumber \\
\dot{A_2} & = & \Lambda_1 A_2 + \alpha_1 \overline{A}_1 A_3 \\
& & \mbox{}+ (a |A_2|^2+b_0 |A_1|^2+b_1|A_3|^2)A_1, \nonumber \\
\dot{A_3} & = & \Lambda_2 A_3 + \alpha_2 A_1 A_2 \\
& & \mbox{} + (b_2 |A_1|^2+b_2 |A_2|^2+b_3|A_3|^2)A_3, \nonumber
\end{eqnarray}
\end{subequations}
to cubic order.  The dot represents differentiation with respect to a 
slow time scale. All coefficients are real.

Because $A_1$ and $A_2$ are neutrally stable modes and $A_3$ is linearly  
damped (\emph{i.e.}, $\Lambda_1=0$ and $\Lambda_2 < 0$), a center manifold reduction 
can be used to eliminate $A_3$. We find
\begin{equation}
\label{eq:cm}
A_3 = -\frac{\alpha_2}{\Lambda_2} A_1 A_2 + \ldots,
\end{equation}
in a neighborhood of the origin.  The (unfolded) bifurcation problem, to 
cubic order, becomes
\begin{subequations}
\label{eq:rhombic-landau}
\begin{eqnarray}
\dot{A}_1 & = & \Lambda_1 A_1+a |A_1|^2A_1 + b(\theta_{\rm res})|A_2|^2A_1, \\
\dot{A}_2 & = & \Lambda_1 A_2+a |A_2|^2A_2 + b(\theta_{\rm res})|A_1|^2A_2,
\end{eqnarray}
\end{subequations}
where
\begin{equation}
b(\theta_{\rm res}) = b_0 + b_{\rm res}, \quad 
b_{\rm res} = -\frac{\alpha_1 \alpha_2}{\Lambda_2}.
\label{eq:bthetar}
\end{equation}
The coefficient $b(\theta)$ is the cross-coupling coefficient for SW 
oriented at an angle $\theta$ relative to each other and, above, it is 
evaluated at the angle of spatial resonance $\theta=\theta_{\rm res}$. 
The resonant contribution $b_{\rm res}$ arises from the presence of the 
damped $k_d$ mode.   

The resonant angle $\theta_{\rm res}$ ranges from 0 to $\pi$ as $k_d$ 
varies from $2k_c$ to 0.  When $k_d$ is such that the natural frequency 
$\Omega(k_d)$ of the damped mode equals (or is nearly equal to) one of 
the special values that promotes a strong nonlinear coupling (as determined in 
Section~\ref{sec:symmetry}) the contribution $b_{\rm res}$ to $b(\theta_{\rm res})$ 
can be significant.  This typically happens when $\alpha_1$ and $\alpha_2$ become  
large in magnitude, and/or when $\Lambda_2$ becomes small in magnitude.  
The resonant contribution will then have a major effect on the stability of 
associated patterns.

\begin{figure}
\centerline{\resizebox{1.6in}{!}{\includegraphics{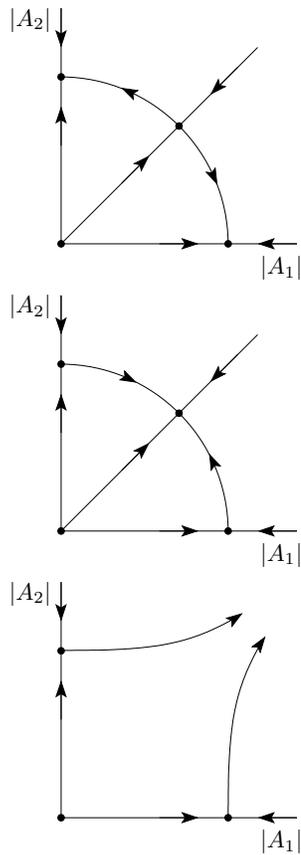}}} 
\caption{Three qualitatively different phase portraits corresponding to 
(\ref{eq:rhombic-landau}) with $a<0$, $\Lambda_1>0$. Top: $b<a$. Middle: 
$a<b<-a$. Bottom: $b>-a$.}
\label{fig:pp}
\end{figure}

Consider further the system~(\ref{eq:rhombic-landau}) which has as 
steady-state solutions the trivial state $|A_1|=|A_2|=0$, the symmetry-related 
``striped'' states $|A_1|>0,|A_2|=0$ and $|A_2|>0,|A_1|=0$, and the ``rhombic'' 
mixed-mode solution $|A_1|=|A_2|$.  We assume that $a < 0$, so that the 
bifurcation to the striped state is supercritical.  A straightforward analysis 
yields the following stability results summarized by Figure~\ref{fig:pp}.   
%%%%%%%%%%%%%%%%%%%%%
%%%%%%%%%%%%%%%%%%%%%
For $b$ sufficiently negative, \emph{i.e.}, $b=b_0+b_{\rm res} < a$, the (supercritical) 
branch of rhombic states with angle $\theta_{\rm res}$ is unstable.  If $b$ is increased 
such that $|b|<|a|$ (typically due to $b_{\rm res}>0$ balancing $b_0<0$) then the two
modes mutually enhance each other's growth, and the rhombic pattern is stable.  If $b$ is 
increased further (due to an even larger, positive $b_{\rm res}$) such that $b>-a$, then the 
rhombic state bifurcates subcritically. However, with the addition of fifth order 
terms (or higher) it is possible, even likely, that for the subcritical case, the unstable 
mixed-mode branch turns around at a saddle-node bifurcation and creates a branch 
of stable, finite amplitude rhombic states. Thus, we do not want to be unduly limited by 
the form of (\ref{eq:rhombic-landau}). In the initial stages of the pattern 
selection process, when modes on the critical circle are beginning 
to grow and compete, there will surely be an advantage for combinations 
that mutually enhance each other's growth.   For these reasons we say that 
triad interactions contributing $b_{\rm res}>0$ are enhancing and those 
giving $b_{\rm res} <0$ are suppressing. 

The above example is just one very basic instance of the importance of 
resonant triads.  In fact, triad resonances have implications far beyond the 
(in)stability of rhombic patterns. They may affect the stability of patterns within 
the framework of any Faraday wave bifurcation problem possessing a subspace 
with dynamics described by (\ref{eq:rhombic-landau}); see, for instance,
\cite{zv1997a,zv1997b,ss1999,sts2000,ts2002,pts2004}. In these cases, the logic is 
the same: $b_{\rm res}>0$ enhances patterns involving the resonant angle 
$\theta_{\rm res}$ while $b_{\rm res} <0$ suppresses them.

The triad interactions discussed in this paper have implications for 
one-dimensional waves as well. In this case, with $\vec{k}_1=\vec{k}_2$, 
the resonance condition (\ref{eq:resonance-condition}) becomes simply
\begin{equation}
2k_c = k_d \,, \label{eq:resonance-1d}
\end{equation}
which is the familiar $\ratio{1}{2}$ spatial interaction.  When the natural 
frequencies of the two waves are such that a strong nonlinear coupling is 
allowed (as we detail in Section~\ref{sec:symmetry}) we expect additional 
contributions to the cubic self-interaction coefficient $a$ in the SW equation
\begin{equation}
\label{eq:1dsw} \frac{d A_1}{dT} = \Lambda_1 A_1+a|A_1|^2A_1 \, ,
\end{equation}
which is simply (\ref{eq:rhombic-landau}) restricted to one spatial
dimension. Since there is no spatial angle $\theta$ to tune, we may 
arrange for a resonant situation (\ref{eq:resonance-1d}) by varying parameters in the dispersion 
relation, as in \cite{ss1999,ts2002}.

\section{Symmetry calculations \label{sec:symmetry}}

We use the approach developed in \cite{ps2002,ps2003,pts2004} to determine 
how the resonant contribution $b_{\rm res}$ to the cross coupling 
coefficient $b(\theta_{\rm res})$ in (\ref{eq:rhombic-landau}) depends 
on the forcing function (\ref{eq:f(t)}). We consider a system of six 
traveling wave (TW) modes (see also \cite{ts2002}) having the same wave 
vectors $\vec{k}_j$ as the three SW modes described in 
Section~\ref{sec:background}.  It is advantageous to consider TW first 
because the action of the temporal symmetries on TW is simple while on SW  
it is not. In this way, we make full use of the temporal symmetry and 
Hamiltonian structure before reducing the TW equations to the desired SW 
equations by means of a center manifold reduction.
We thus expand the fluid surface height $h(\vec{x},t)$, 
$\vec{x}\in \mathbb{R}^2$ as
\begin{equation}
\label{eq:twmodes1}
h(\vec{x},t) = \sum_{j=1}^{3} \sum_{\pm} Z^{\pm}_j(t) \exp{i(\vec{k}_j  
\cdot \vec{x} \pm \varpi_j t)} + c.c.,
\end{equation}
where $Z_j^{\pm}$ are the slowly varying amplitudes and 
\begin{equation}
\label{eq:freqvals}
\varpi_1=\varpi_2=m/2,\quad \varpi_3 = \Omega.
\end{equation}
Spatial and temporal symmetries constrain the equations for the 
evolution of  $Z_j^{\pm}$, as we now detail.

\subsection{Spatial symmetries \label{sec:spatial}}

Spatial translation symmetry acts on the TW amplitudes $Z_j^{\pm}$ 
as (\emph{cf.} Eq.~\ref{eq:SWspatialtrans})
\begin{gather}
\label{eq:spatialtrans}
T_{\Theta}: Z_j^{\pm} \rightarrow Z_j^{\pm} \exp{i\theta_j},  
\\*
\Theta = (\theta_1,\theta_2),\quad \theta_1,\theta_2 \in [0,2\pi),  
\quad \theta_3 = \theta_1+\theta_2. \nonumber
\end{gather}
A reflection about $\vec{k}_3$ acts as (\emph{cf.} Eq.~\ref{eq:SWspatialreflec2})
\begin{equation}
\mathcal{\kappa}: Z_1^{\pm} \leftrightarrow Z_2^{\pm},   
\label{eq:spatialreflec2}
\end{equation}
and a rotation by $\pi$ induces (\emph{cf.} Eq.~\ref{eq:SWspatialreflec1})
\begin{equation}
\mathcal{R}: Z_j^{\pm} \rightarrow \conjugate{Z}_j^{\mp}  
\label{eq:spatialreflec1}.
\end{equation}

We enforce equivariance under (\ref{eq:spatialtrans})--(\ref{eq:spatialreflec1}) 
to obtain the form of the TW amplitude equations to quadratic order. This 
truncation is sufficient to determine the leading order resonant contribution 
$b_{\rm res}$ to $b(\theta_{\rm res})$ in (\ref{eq:rhombic-landau}).  We have
\begin{subequations}
\label{eq:qtwae}
\begin{eqnarray}
\dot{Z}_1^+ & =  & L_1 Z_1^+ + L_2 Z_1^- + Q_1 \conjugate{Z}_2^+ Z_3^+\\
& & \mbox{} +  Q_2 \conjugate{Z}_2^+ Z_3^- +  Q_3 \conjugate{Z}_2^-  
Z_3^+ +  Q_4 \conjugate{Z}_2^- Z_3^-, \nonumber \\
\dot{Z}_3^+ & =  & L_3 Z_3^+ +L_4 Z_3^- +  Q_5 Z_1^+Z_2^+  \\
& & \mbox{}+   Q_6 Z_1^+Z_2^- + Q_6 Z_1^- Z_2^+ +  Q_7 Z_1^- Z_2^-, 
\nonumber
\end{eqnarray}
\end{subequations}
where the remaining four equations follow from the application of  
(\ref{eq:spatialreflec2}) and~(\ref{eq:spatialreflec1}).

We now apply a standard reduction procedure to (\ref{eq:qtwae}) and compare 
this result with the SW equations~(\ref{eq:rhombic-landau}).  To facilitate 
the subsequent calculations we first introduce a phase shift to the amplitudes: 
\begin{equation}
Z_{1,2}^{\pm} \rightarrow Z_{1,2}^{\pm} \exp{\pm i\varphi/2},
\end{equation}
where
\begin{equation}
\label{eq:phi}
\varphi =  \varphi_2 - \varphi_1 + \pi,
\end{equation}
with $\varphi_{1,2}$ denoting the phases of the coefficients $L_1$ and 
$L_2$  (\emph{i.e.}, $L_{1,2}=|L_{1,2}|\exp{i\varphi_{1,2}}$).  The TW 
equations (\ref{eq:qtwae}) may be compactly written in the form
\begin{equation}
\dot{\vec{Z}} = \mat{L} \vec{Z} + \vec{N}(\vec{Z}),
\end{equation}
where $\vec{Z} = (Z_1^+, Z_1^-, Z_2^+, Z_2^-, Z_3^+, Z_3^-)^T$.

The bifurcation to SW occurs when $|L_2|=|L_1|$.   As we will see in the  
next section, $|L_2| \sim |f_m|$, so this bifurcation condition serves to 
define the critical amplitude of the dominant forcing component $|f_m|$.  
The critical eigenvectors are $\vec{v}_1 = (1,1,0,0,0,0)^T$ and 
$\vec{v}_2 =  (0,0,1,1,0,0)^T$.  We use a multi-scale perturbation 
calculation to accomplish the reduction to SW, writing
\begin{subequations}
\begin{eqnarray}
\vec{Z} & = & \eta\,(A_1 \vec{v}_1 + A_2 \vec{v}_2) + \eta^2\vec{Z}_2 
+ \ldots\,, \\
|f_m| & = & |f_m|^{crit} + \eta^2 |f_{2}| + \ldots \,,  \\
\frac{d}{dt} & = & \eta^2 \frac{\partial}{\partial T_2} + \ldots \,,
\end{eqnarray}
\end{subequations}
where $\eta \ll 1$ is a small bookkeeping parameter and $A_{1,2}$ are  
the time-dependent SW amplitudes.  At $\order(\eta)$ the linear problem  
is recovered.  At $\order(\eta^2)$ $\vec{Z}_2$ is determined.  At  
$\order(\eta^3)$ a solvability condition yields equations for the slow 
variation of the SW amplitudes:
\begin{subequations}
\label{eq:A1A2}
\begin{eqnarray}
&\dot{A}_1 = \Lambda_1 A_1 + b_{\rm res} |A_2|^2 A_1,\\
&\dot{A}_2 = \Lambda_1 A_2 + b_{\rm res} |A_1|^2 A_2.
\end{eqnarray}
\end{subequations}
The coefficients $a$ and $b_0$ in (\ref{eq:rhombic-landau}) do not appear 
above because the cubic order terms were omitted in (\ref{eq:qtwae}).  
For the purposes of this paper, we need only point out that the ``nonresonant'' 
coefficients $a$ and $b_0$ are both $\order(\gamma)$ 
\cite{ps2002,ts2002,ps2003} (recall that $\gamma$ is a dimensionless 
measure of the damping).  The resonant contribution is given by
\begin{equation}
\label{eq:bres}
b_{\rm res} = \frac{\re\{\conjugate{L}_1S\}}{\re\{L_1\}},
\end{equation}
where
\begin{equation}
S = Q_1W\exp{-i\varphi} + Q_2 \conjugate{W}\exp{-i\varphi} + Q_3 W + Q_4  
\conjugate{W},
\end{equation}
with
\begin{eqnarray}
W & = & (L_4 \conjugate{U}-\conjugate{L}_3 U)/(|L_3|^2-|L_4|^2), \nonumber \\
U & = & \exp{i\varphi}Q_5 + 2Q_6 + \exp{-i\varphi}Q_7,
\end{eqnarray}
and $\varphi$ defined by (\ref{eq:phi}).  Our analysis applies when $|L_3|>|L_4|$, \emph{i.e.} when the $\vec{k_3}$ mode is linearly damped.

\subsection{Temporal symmetries \label{sec:temporal}}

Temporal symmetries constrain the coefficients $L_1, \ldots, L_4$ and 
$Q_\ell$, $\ell = 1, \ldots, 7$ in (\ref{eq:qtwae}).  In the absence of damping 
and forcing, the problem has an exact time translation symmetry
\begin{equation}
\label{eq:plaintimetrans}
T_{\Delta t}: Z_j^{\pm} \rightarrow Z_{j}^{\pm} \exp{\pm i \varpi_j \Delta t},
\end{equation}
with $\varpi_j$ given by (\ref{eq:freqvals}), and an exact time  
reversal symmetry
\begin{equation}
\label{eq:plaintimerev}
\sigma : t \rightarrow -t,\quad Z_j^{\pm} \rightarrow Z_j^{\mp}.
\end{equation}
In the presence of finite damping and forcing, these temporal symmetries are 
broken.  Nonetheless, they can be recast as unbroken {\it parameter} 
symmetries by allowing an appropriate transformation of the forcing 
parameters $f_u$ and the damping $\gamma$.   With this generalization the 
time translation symmetry (\ref{eq:plaintimetrans}) becomes
\begin{equation}
\label{eq:timetrans}
T_{\Delta t}: Z_j^{\pm} \rightarrow Z_{j}^{\pm} \exp{\pm i  
\varpi_j \Delta t}, \quad f_u \rightarrow f_u \exp{iu \Delta t},
\end{equation}
and the time reversal symmetry (\ref{eq:plaintimerev}) becomes
\begin{equation}
\label{eq:timerev}
\sigma:(t,\gamma) \rightarrow -(t,\gamma) ,\quad Z_j^{\pm} \rightarrow  
Z_j^{\mp},\quad f_u \rightarrow \conjugate{f}_u.
\end{equation}

The damping and forcing are both assumed to be small, and are of the same  
order, \emph{i.e.}, $|f_u| \sim \gamma \ll 1$.   A Taylor  
expansion of the coefficients $L_1, \ldots, L_4$ and $Q_\ell$, consistent 
with (\ref{eq:timetrans}) and (\ref{eq:timerev}), leads to
\begin{subequations}
\label{eq:coeff-expansions}
\begin{eqnarray}
L_1 & = & -\upsilon_r \gamma , \\
L_2 & = & -i \lambda_i f_m , \\
L_3 & = & -\varrho_r \gamma ,\\
L_4 & = &  -i \mu_i F_{2\Omega}, \\
Q_\ell & = & i q_\ell F_\ell , \label{eq:ql}
\end{eqnarray}
\end{subequations}
where only the leading order terms have been kept.  
The expansion coefficients are all real, and $\upsilon_r, \varrho_r > 0$ 
since they correspond to damping terms.
The factor of $f_m$ in the expansion of $L_2$ reflects the fact that the 
critical modes respond parametrically to the dominant component $f_m$. 
The factor $F_{2\Omega}$ in the expansion of $L_4$ represents an analogous 
parametric forcing term for the damped mode composed of products of 
the $f_u$ and $\bar{f}_u$ whose frequencies sum to $2\Omega$.  When $2\Omega$ 
forcing is present in (\ref{eq:f(t)}), then, to lowest order, 
$F_{2\Omega} = f_{2\Omega}$; otherwise $L_4=0$ at $\order(\gamma)$.    

The $F_\ell$ in (\ref{eq:ql}), in accordance with (\ref{eq:timetrans}), must contain products 
of the $f_u$ (and $\bar{f}_u$) whose frequencies are such that 
$(Q_1,\conjugate{Q}_5)\exp{i(\Omega-m) t}$, $(Q_2,Q_7)\exp{-i(m+\Omega) t}$, 
and $(Q_3,\conjugate{Q}_4,\conjugate{Q}_6)\exp{i\Omega t}$ are time 
translation-invariant quantities.   Since we are interested in understanding   
when the effect of resonant triads is significant, we focus on those cases  
where $b_{\rm res}$ is $\order(\gamma)$ or larger; this requires that 
one or more of the quadratic coefficients $Q_\ell$ is $\order(\gamma)$ 
or larger.  A straightforward calculation shows that this scaling can 
occur only if $\Omega \in \{m,2m,n,m \pm m,n-m\}$, $\Omega > 0$, for some 
frequency $n$, and we henceforth restrict attention to these cases.   Note 
that, since there are many frequencies in $f(\tau)$, these sets of 
relevant $\Omega$ values can overlap. For instance, an ``$m-n$ mode'' is also 
a ``$p-m$ mode'' if $n+p=2m$, $n<m$.  An important (and somewhat surprising) 
result of our symmetry calculation is that the contribution $b_{\rm res}$ 
arising from a given damped mode with frequency $\Omega$ involves (at 
leading order) a maximum of five frequencies: the dominant frequency $m$, 
up to three other frequencies appearing at $\order(\gamma)$ in the three 
sets of couplings coefficients $(Q_1,\conjugate{Q}_5)$, $(Q_2,Q_7)$, and 
$(Q_3,\conjugate{Q}_4,\conjugate{Q}_6)$, and potentially one more 
frequency, $2\Omega$, that parametrically forces the damped mode at 
$\order(\gamma)$, thus making $L_4$ nonzero at leading order. The effect on the triad 
interaction of any additional forcing components in (\ref{eq:f(t)}) will 
be $\order(\gamma^2)$ or smaller.

\subsection{Results \label{sec:results}}

We combine the results of Sections \ref{sec:spatial} and
\ref{sec:temporal} to obtain leading order expressions for $b_{\rm res}$ 
in (\ref{eq:rhombic-landau}) with explicit dependence on the damping 
$\gamma$, the forcing amplitudes $|f_u|$ and the forcing phases $\phi_u$.  
For each $\Omega$ there are a finite number of qualitatively distinct cases to 
consider.  These are distinguished by the number of relevant frequencies 
involved (up to five) and the manner in which they enter the problem (through 
$L_4$ and the $Q_\ell$).  Having chosen one of these, we substitute 
the corresponding expressions for $Q_\ell$, and the expansions for the 
remaining TW coefficients shown in (\ref{eq:coeff-expansions}), into 
equation~(\ref{eq:bres}) for $b_{\rm res}$.   The results are
summarized in Table~\ref{tab:main-results}, and will be discussed in
the next section.

To make this table of results manageable, we make use of the following 
definitions:
\begin{subequations}
\label{eq:shorthand1}
\begin{eqnarray}
\alpha_1 &=& q_{1}q_{5}, \\
\alpha_2 &=& q_{2}q_{7}, \\
\alpha_3 &=& 2q_{6}(q_{3}-q_{4}),\\
\alpha_4 &=& q_{1}q_{7}-q_{2}q_{5}, \\
\alpha_5 &=& \left\{2q_{1}q_{6}+q_{5}(q_{3}-q_{4}) \right\} 
\lambda_i /|\lambda_i|, \\
\alpha_6 &=& \left\{2q_{2}q_{6}-q_{7}(q_{3}-q_{4}) \right\}
\lambda_i /|\lambda_i|,
\end{eqnarray}
\end{subequations}
and
\begin{subequations}
\label{eq:shorthand2}
\begin{gather}
P_{2\Omega}(\Phi)=\frac{|L_3|+\mu_i|f_{2\Omega}|\sin\Phi}{|L_3|^2-|\mu_i
f_{2\Omega}|^2}, \label{eq:P}\\
R_{2\Omega}(\Phi_1,\Phi_2)=\frac{|L_3|\sin\Phi_1+\mu_i|f_{2\Omega}|
\cos\Phi_2}{|L_3|^2-|\mu_i f_{2\Omega}|^2}. \label{eq:R}
\end{gather}
\end{subequations}
In the above, the $q_j$ and $\lambda_i$ are defined by (\ref{eq:coeff-expansions}). 
The relevant phases $\Phi,\Phi_1,\Phi_2$ appear in Table \ref{tab:main-results}.

\begingroup
\begin{table*}
\caption{\label{tab:main-results} Leading resonant contribution $b_{\rm res}$ 
to $b$ in (\ref{eq:rhombic-landau}) for the most important damped 
modes.  For a damped mode with frequency $\Omega$, there are at most 
five forcing frequencies $m,n,p,q,r$ which affect $b_{\rm res}$.  Here, 
$m,n,p,q,r,\Omega > 0$ and  $x\in\mathbb{Z}^+$.  Each expression for 
$(m,n,p,q,r)$, given $\Omega$, is excluded from those of entries further down 
the table, in which additional relationships hold.  Dots indicate an arbitrary 
commensurate frequency, if present, which does not affect $b_{\rm res}$ at lowest 
order.  Entries whose listed frequencies have a common factor 
(\emph{e.g.}, those with $x$) are assumed to be part of a forcing 
function with other, relatively prime, frequencies.  For $\star$ the $\pm$ 
follows $\sgn(m-n)$.  See (\ref{eq:shorthand1}) and (\ref{eq:shorthand2}) 
for definitions of $\alpha_1, \ldots, \alpha_6$, $P_{2\Omega}$, and 
$R_{2\Omega}$ used below.  Certain entries are reproduced from \cite{pts2004}; 
the cases that involve more than three forcing frequencies are new.}
\begin{tabular*}{\textwidth}{@{\extracolsep{\fill}}lccl}
\hline \hline
$(m,n,p,q,r)$ & $\Omega$ & Leading resonant contribution $b_{\rm res}$  
& relevant phase(s) \\ \hline \hline
$(m,\cdot, \cdot, \cdot, \cdot)$ &$m$ &$-\alpha_1/|L_3|$ & \\
$(m,\cdot, \cdot, \cdot, \cdot)$ &$2m$ &$-\alpha_1|f_m|^2/|L_3|$ &  
\\
$(m,n, \cdot, \cdot, \cdot)$ &$n$ &$-\alpha_3|f_n|^2/|L_3|$ & \\
$(m,n, \cdot, \cdot, \cdot)$ &$m \pm n$ &$-\alpha_1|f_n|^2/|L_3|$ &  
\\
$(m,n, \cdot, \cdot, \cdot)$ &$n-m$ &$\alpha_2|f_n|^2/|L_3|$ & \\  
\hline \hline
$(m,2m, \cdot, \cdot, \cdot)$ &$m$ &$-\alpha_1P_n(\Phi)$  
&$\Phi=\phi_n \!-  2\phi_m$\\
$(3x,2x, \cdot, \cdot, \cdot)$ &$x$ &$-\alpha_1|f_n|^2P_n(\Phi)$  
&$\Phi=3\phi_n \!- 2\phi_m$ \\
$(m,4m,\cdot, \cdot, \cdot)$ &$2m$ &$-\alpha_1|f_m|^2P_n(\Phi)$  
&$\Phi=\phi_n \!- 4\phi_m$ \\
$(m,n,2n, \cdot, \cdot)$ &$n$ &$-\alpha_3|f_n|^2P_p(\Phi)$  
&$\Phi=2\phi_n \!- \phi_p$ \\
$(m,n,2m \pm 2n, \cdot, \cdot)$ &$m \pm n$ &$-\alpha_1|f_n|^2P_p(\Phi)$  
&$\Phi=\phi_p \!- 2\phi_m \!\mp 2\phi_n$ \\
$(m,n,2n-2m, \cdot, \cdot)$ &$n-m$ &$\alpha_2|f_n|^2P_p(\Phi)$  
&$\Phi=\phi_p \!+ 2\phi_m \!- 2\phi_n 
$\\ \hline \hline
$(m,2m,\cdot, \cdot, \cdot)$ &$2m$  
&$(-\alpha_1|f_m|^2-\alpha_3|f_n|^2+\alpha_5|f_m||f_n|\sin\Phi)/ 
|L_3|$ &$\Phi=\phi_n  \!- 2\phi_m $ \\
$(m,3m,\cdot, \cdot, \cdot)$ &$2m$  
&$(-\alpha_1|f_m|^2+\alpha_2|f_n|^2+\alpha_4|f_m||f_n|\cos\Phi)/ 
|L_3|$ &$\Phi=\phi_n \!- 3 \phi_m$ \\
$(m,n,|m-n|, \cdot, \cdot)$ &$n$ &$(-\alpha_1|f_p|^2- \alpha_3|f_n|^2+  
\alpha_5|f_n||f_p|\sin\Phi)/|L_3|$ &$\Phi=\phi_n \!- \phi_m \!\pm  
\phi_p \,\star$ \\
$(m,n,m+n, \cdot, \cdot)$ &$n$ &$(\alpha_2|f_p|^2- \alpha_3|f_n|^2 +  
\alpha_6|f_n||f_p|\sin\Phi)/|L_3|$ &$\Phi=\phi_m \!+ \phi_n \!- \phi_p$ \\
$(m,n,2m \pm n, \cdot, \cdot)$ &$m \pm n$ &$(\alpha_2|f_p|^2  
-\alpha_1|f_n|^2 + \alpha_4|f_n||f_p|\cos\Phi)/|L_3|$ &$\Phi=2  
\phi_m \!- \phi_p \!\pm \phi_n$ 
\\ \hline \hline
$(3x,x,2x, \cdot, \cdot)$ &$x$ &$-\alpha_1|f_p|^2 P_p(\Phi_1 \!- \Phi_2)  
- \alpha_3|f_n|^2 P_p(\Phi_1 \!+ \Phi_2)$ 
& $\Phi_1=\phi_n \!- \phi_m  \!+ \phi_p$ \\
& &$\mbox{}+ \alpha_5|f_n||f_p|R_p(\Phi_1,\Phi_2)$ &$\Phi_2=\phi_m \!+  
\phi_n \!- 2\phi_p$ \\
$(3x,2x,4x, \cdot, \cdot)$ &$x$&$-\alpha_1|f_n|^2 P_n(\Phi_1 \!+ \Phi_2)  
+ \alpha_2|f_p|^2 P_n(\Phi_2 \!- \Phi_1)$ 
&$\Phi_1=\phi_n \!+ \phi_p \!- 2\phi_m$ \\
& &$+ \alpha_4|f_n||f_p|R_n(\Phi_1 \!- 90^\circ,\Phi_2 \!+ 90^\circ)$  
&$\Phi_2=2\phi_n \!- \phi_p$ \\
$(m,2m,4m, \cdot, \cdot)$ &$2m$ &$-\alpha_1|f_m|^2 P_p(\Phi_1 \!- \Phi_2)  
- \alpha_3|f_n|^2 P_p(\Phi_1 \!+ \Phi_2)$ 
&$\Phi_1=\phi_n \!- 2\phi_m$ \\
& & $\mbox{}+ \alpha_5|f_m||f_n|R_p(\Phi_1,\Phi_2)$ 
&$\Phi_2=\phi_n \!+ 2\phi_m \!- \phi_p$ \\
$(m,3m,4m, \cdot, \cdot)$ &$2m$ &$-\alpha_1|f_m|^2 P_p(\Phi_1 \!- \Phi_2)  
+ \alpha_2|f_n|^2 P_p(\Phi_1 \!+ \Phi_2 \!+ 180^\circ) $ 
&$\Phi_1=\phi_n \!- 3\phi_m$ \\
& &$\mbox{}+ \alpha_4|f_m||f_n|R_p(\Phi_1 \!+ 90^\circ,\Phi_2 \!+ 90^\circ)$  
&$\Phi_2=\phi_m \!+ \phi_n \!- \phi_p$ \\
$(m,n,|m-n|,2n,\cdot)$ & $n$ & $-\alpha_1|f_p|^2 P_q(2\Phi_1 \!- \Phi_2) -  
\alpha_3|f_n|^2 P_q(\Phi_2)$ 
& $\Phi_1=\phi_n \!- \phi_m \!\pm \phi_p\,\star$ \\
& &$\mbox{}+ \alpha_5|f_n||f_p|R_q(\Phi_1,\Phi_2 \!- \Phi_1)$ 
&$\Phi_2=2\phi_n \!- \phi_q$ \\
$(m,n,m+n,2n,\cdot)$ & $n$ & $\alpha_2|f_p|^2 P_q(2\Phi_1 \!- \Phi_2) -  
\alpha_3|f_n|^2 P_q(\Phi_2)$ 
& $\Phi_1=\phi_m \!+ \phi_n \!- \phi_p$ \\
& &$\mbox{}+ \alpha_6|f_n||f_p|R_q(\Phi_1,\Phi_1 \!- \Phi_2)$ 
&$\Phi_2=2 \phi_n \!- \phi_q$ \\
$(m,n,2m \pm n,2m \pm 2n,\cdot)$ & $m \pm n$ &$-\alpha_1|f_n|^2 
P_q(\Phi_2 \!- 2 \Phi_1) + \alpha_2 |f_p|^2P_q(\Phi_2) $ 
& $\Phi_1 = 2\phi_m \!\pm \phi_n \!- \phi_p$\\
& & $\mbox{}+ \alpha_4 |f_n||f_p|R_q(\Phi_1 \!+ 90^\circ,\Phi_2 \!- \Phi_1 \!- 90^\circ)$ 
&  $\Phi_2 = 2\phi_m \!- 2\phi_p \!+ \phi_q$ \\ 
\hline \hline
$(m,2m,3m, \cdot, \cdot)$ & $2m$ & $\{(\alpha_2|f_p|^2 -  
\alpha_1|f_m|^2 - \alpha_3|f_n|^2 + \alpha_4|f_m||f_p|\cos\Phi_1$  
&$\Phi_1=\phi_p \!- 3\phi_m$ \\
& &$\mbox{}+\alpha_5|f_m||f_n|\sin\Phi_2 +  
\alpha_6|f_n||f_p|\sin(\Phi_2 \!- \Phi_1)\}/|L_3|$ 
& $\Phi_2=\phi_n - 2\phi_m$ \\
$(m,n,m+n,|m-n|,\cdot)$ & $n$ & $\{(\alpha_2|f_p|^2 - \alpha_1|f_q|^2 -  
\alpha_3|f_n|^2 + \alpha_4|f_p||f_q|\cos\Phi_1$ 
&$\Phi_1=\phi_p \!- 2\phi_m \!\pm \phi_q \,\star$ \\
& &$\mbox{}+\alpha_5|f_n||f_q|\sin(\Phi_1 \!+ \Phi_2) +  
\alpha_6|f_n||f_p|\sin\Phi_2 \}/|L_3|$ 
& $\Phi_2=\phi_m \!+ \phi_n \!- \phi_p$ \\ 
\hline \hline
$(3x,x,4x,2x,\cdot)$ & $x$ & $-\alpha_1 |f_q|^2  
P_q(\Phi_2 \!- \Phi_1)+\alpha_2 |f_p|^2 P_q(\Phi_1 \!+ \Phi_2)$
&$\Phi_1 = 2 \phi_m  \!- \phi_p \!- \phi_q$\\
& &$\mbox{}-\alpha_3|f_n|^2  P_q(2\Phi_3 \!- \Phi_1 \!- \Phi_2)+
\alpha_4|f_p||f_q|R_q(\Phi_1 \!+ 90^\circ, \Phi_2 \!- 90^\circ) $
&$\Phi_2 = 2\phi_q \!- \phi_p$\\
& &  
$\mbox{}+\alpha_5|f_n||f_q|R_q(\Phi_3 \!- \Phi_1,\Phi_3 \!- \Phi_2)
+\alpha_6|f_n||f_p|R_q(\Phi_3,\Phi_1 \!+\Phi_2 \!- \Phi_3)$
& $\Phi_3 = \phi_m  \!+ \phi_n \!- \phi_p$ \\
$(m,2m,3m,4m,\cdot)$ & $2m$ & $-\alpha_1 |f_m|^2  
P_q(-\Phi_1 \!- \Phi_2)+\alpha_2 |f_p|^2 P_q(\Phi_2 \!- \Phi_1)$
&$\Phi_1 = \phi_m \!+ \phi_p \!- \phi_q$\\
& &$\mbox{}-\alpha_3|f_n|^2 P_q(2\Phi_3 \!- \Phi_1 \!- \Phi_2)
+\alpha_4|f_m||f_p|R_q(\Phi_2 \!+90^\circ,\Phi_1 \!+ 90^\circ) $
&$\Phi_2 = 3 \phi_m \!+ \phi_p$\\
& &  $\mbox{}+\alpha_5|f_m||f_n|R_q(\Phi_3 \!- \Phi_1 \!- \Phi_2,\Phi_3)
+\alpha_6|f_n||f_p|R_q(\Phi_3 \!- \Phi_1,\Phi_3 \!- \Phi_2)$ 
& $\Phi_3 = 2\phi_m \!+ \phi_n \!- \phi_q$ \\
$(m,n,m+n,|m-n|,2n)$ & $n$ & $-\alpha_1 |f_q|^2 P_r(\Phi_2 \!- \Phi_1)+
\alpha_2 |f_p|^2 P_r(\Phi_1 \!+ \Phi_2)$
&$\Phi_1 = 2\phi_m \!- \phi_p \!\mp \phi_q\,\star$\\
& & $\mbox{}-\alpha_3|f_n|^2P_r(2\Phi_3 \!- \Phi_1 \!- \Phi_2)+
\alpha_4|f_p||f_q|R_r(\Phi_1 \!+ 90^\circ,\Phi_2 \!- 90^\circ) $
& $\Phi_2 = \phi_r \!- \phi_p \!\pm \phi_q\,\star$\\
& &  
$\mbox{}+ \alpha_5|f_n||f_q|R_r(\Phi_3 \!- \Phi_1,\Phi_2 \!- \Phi_3)
+\alpha_6|f_n||f_p|R_r(\Phi_3,\Phi_2 \!+ \Phi_3 \!- \Phi_1)$ 
&$\Phi_3 = \phi_m \!+ \phi_n \!- \phi_p$ \\  
\hline \hline
\end{tabular*}
\end{table*}
\endgroup

\subsection{Hamiltonian structure \label{sec:hamiltonian}}

We now discuss the implications of Hamiltonian structure in the undamped problem 
(see \cite{zv1997a,m1991,z1968,m1977,b1974,la1997,r1992}).  This is a stronger 
assumption than that of time reversal symmetry~(\ref{eq:timetrans}) alone.   
We suppose, as in \cite{ps2002,ps2003,pts2004}, that the undamped TW 
equations~(\ref{eq:qtwae}) can be derived from a Hamiltonian $\mathcal{H}$.
Because the amplitudes $Z_j^\pm$ and $\bar{Z}_j^\pm$ need not themselves be  
canonically conjugate Hamiltonian variables, we write Hamilton's equations in the 
generalized form  
\begin{equation}
\label{eq:hamiltonequations}
\dot{Z}_j^{\pm} = \mp \frac{1}{r_j^2} \frac{\partial \mathcal{H}}{\partial
\conjugate{Z}_j^{\pm}}, \qquad r_1=r_2, \quad r_j \in \mathbb{R}.
\end{equation}
This takes account of scaling transformations like  
$(Z_1^{\pm},Z_2^{\pm}) \rightarrow r_1 (Z_1^{\pm},Z_2^{\pm})$,  
$Z_3^{\pm} \rightarrow r_3 Z_3^{\pm}$ that preserve the Hamiltonian 
character of the dynamics, and are needed to relate the underlying canonical 
variables to $Z_j^\pm$ and $\bar{Z}_j^\pm$ in (\ref{eq:qtwae}).    For inviscid 
Faraday waves the surface height $h$ and the surface velocity potential are the 
underlying canonical variables (see, \emph{e.g.}, \cite{z1968,m1977}).    Using this 
fact we find that, to leading order,  $r_1^2=r_2^2=m/(2k_c)$ and $r_3^2=\Omega/k_d$ 
are appropriate prefactors  in~(\ref{eq:hamiltonequations}) (see \cite{k1994} where a 
similar factor arises in the corresponding canonical transformation).     

Requiring that $\mathcal{H}$ be a real-valued function, invariant under the
symmetries (\ref{eq:spatialtrans})--(\ref{eq:spatialreflec1}), (\ref{eq:timetrans})
and (\ref{eq:timerev}), we find that the equations of motion~(\ref{eq:hamiltonequations}) 
are equivalent to (\ref{eq:qtwae}) only if $q_1 = r q_5$, $q_2 = r q_7$, and 
$q_3=q_4 = r q_6$ with $r = r_3^2/r_1^2$.  These conditions imply, for the results 
in Table~\ref{tab:main-results}, that
\begin{equation}
\label{eq:ham}
\alpha_1 > 0,\quad \alpha_2 > 0,\quad \alpha_3 = 0, \quad \alpha_4 = 0.
\end{equation}

\section{Discussion \label{sec:discussion}}

We now discuss Table~\ref{tab:main-results} in some detail, highlighting 
the most important features of the results collected there.  We then 
investigate the range of validity of these results, which were derived 
under the assumption of weak damping $\gamma$. To do this, we introduce 
the Zhang-Vi\~{n}als Faraday wave equations and use them to perform 
explicit numerical calculations that demonstrate the range of $\gamma$ 
for which the symmetry-based results provide an accurate prediction.

\subsection{Highlights of results \label{sec:highlights}}

Some general comments on the organization of Table~\ref{tab:main-results} 
are in order.  Note first that there are many cases which do not need 
to be listed because they can be obtained simply by relabeling the different 
frequencies.  For example, the case $(m,n,p,q,\cdot \,;\Omega)=
(m,n,2m+n,m+n,\cdot \,;m+n)$ is equivalent to the case (fourth up from the bottom 
in Table~\ref{tab:main-results}) $(m,n,p,q,\cdot \,;\Omega)=
(m,n,m+n,n-m,\cdot \,;n)$ with $n \leftrightarrow q$. 

There are six groupings in the table.  The first shows the five  
important damped modes and their contribution to $b_{\rm res}$ when  
there is only one type of coupling at ${\cal O}(\gamma)$ or lower 
and no parametric forcing $f_{2\Omega}$.  In these cases there is no  
(leading order) dependence on the forcing phases $\phi_u$.  In the second  
section the same damped modes have been parametrically forced.  The  
factor $1/|L_3|$ is then replaced by $P_{2\Omega} (\Phi)$ of 
(\ref{eq:P}).  This is a strictly positive oscillatory function  
($|L_3| > |\mu_if_{2\Omega}|$ for damped modes) with extrema at 
$\Phi = \pm 90^\circ$.  The third and fourth sections are analogous to 
the first and second, but with two types of coupling rather than 
one -- similarly for the fifth and sixth sections, but with all three 
possible quadratic couplings (\emph{i.e.}, all $Q_\ell$ are linear in 
the $f_u$).

Two of the damped modes appearing in the table warrant special mention.
The $\Omega = m$ mode stands out because its influence is especially
strong.  For this mode, the largest quadratic terms in (\ref{eq:qtwae})
are $\order(1)$, and the resulting contribution $b_{\rm res}$ is
$\order(\gamma^{-1})$.  In contrast, for all of the other damped modes,
the strongest quadratic couplings take place at $\order(\gamma)$ and
lead to $b_{\rm res}$ of $\order(\gamma)$; these $\order(\gamma)$ 
contributions are of the same order as $a$ and $b_0$ in 
(\ref{eq:rhombic-landau}), but can still have significant effects on 
pattern seelction, as demonstrated in Section~\ref{sec:applications}.

The second special case is the $\Omega = 2m$ mode.  Although this mode 
satisfies all the necessary temporal constraints to make a significant 
contribution $b_{\rm  res}$, it cannot enter into resonant triad interactions 
with the critical modes because its wave number is too large, \emph{i.e.}, 
$k_d > 2k_c$ and (\ref{eq:resonance-condition}) cannot be satisfied; one 
can estimate the relevant wave numbers from the inviscid fluid dispersion relation 
(see \cite{ts2002} and Section~\ref{sec:zv}).  However, this mode may 
have relevance for other systems such as ferrofluids in a magnetic field 
where the dispersion relation is nonmonotonic \cite{mgr1996}, and hence 
we have kept it in the table.

A key result of Table~\ref{tab:main-results} is the important role played 
by the relative phases $\phi_u$ in the forcing function~(\ref{eq:f(t)}).   
For all but the most simple cases (in the first section of the table), 
$b_{\rm res}$ depends on combinations of the forcing phases which are 
invariant under the time translation symmetry $T_\tau$ of (\ref{eq:timetrans}); 
one phase is always arbitrary, associated with the choice of origin in time, 
while any physically meaningful phase must be invariant under~(\ref{eq:timetrans}).
This phase dependence provides a very convenient way to tune the strength 
of the nonlinear interactions, as the numerical examples of 
Section~\ref{sec:applications} will demonstrate.

Another important aspect of Table~\ref{tab:main-results} pertains to the
sign of $b_{\rm res}$.  Recall from the discussion of 
Section~\ref{sec:background} that if $b_{\rm res}>0$ interactions involving 
critical modes separated by the angle $\theta_{\rm res}$ will
be enhanced, whereas if $b_{\rm res}<0$ they will be suppressed. 
Relations~(\ref{eq:ham}) mean that for simple couplings (the first two 
sections of Table~\ref{tab:main-results}) the sign of $b_{\rm res}$ is 
determined, and thus one knows which effect (if any) to expect.  In 
particular, the $\Omega=m$,  $\Omega=2m$, and $\Omega = m \pm n$ 
modes are suppressing while the $\Omega=n$ mode is inconsequential.  
The $\Omega=n-m$ mode, in contrast, is enhancing, and thus is of great
interest because it may be used directly as a selection mechanism.  
The effect of this difference frequency mode on pattern selection was 
examined in \cite{ts2002}, and indeed, it is likely responsible for stabilizing the 
superlattice pattern observed in \cite{kpg1998}. We examine 
the difference frequency mode further in Section~\ref{sec:applications}. 

A final noteworthy feature of Table~\ref{tab:main-results} concerns the 
effect of parametrically forcing the damped mode with a frequency $2\Omega$.   
A comparison of the factors $1/|L_3|$, $P_{2\Omega}(\Phi)$, and 
$R_{2\Omega}(\Phi_1,\Phi_2)$ reveals the potential for a small denominator 
in the latter two cases.  The parametric forcing can increase $|b_{\rm res}|$ and amplify 
the effect of the damped mode provided this denominator does not become 
excessively small, which would indicate that the damped mode is nearly 
critical and that the reduction leading to (\ref{eq:A1A2}) is breaking 
down.
This feature will be exploited as well in some of the examples of 
Section~\ref{sec:applications}.

\subsection{Zhang-Vi\~{n}als hydrodynamic equations\label{sec:zv}}

In this subsection, we investigate the range of damping $\gamma$ for 
which our symmetry-based results are valid.  To carry out this 
investigation we perform explicit numerical calculations using the 
Zhang-Vi\~{n}als hydrodynamic equations (introduced below).  In 
particular, we use the method described in  \cite{sts2000} to calculate 
the cross-coupling coefficient $b$ in (\ref{eq:rhombic-landau}) as a 
function of $\theta$, the angle between $\vec{k}_1$ and $\vec{k}_2$ 
in Figure~\ref{fig:resonant_triad}.  It is sufficient to take 
$\theta \in [0,90)$ since $b(180^\circ-\theta)=b(\theta)$.

The Zhang-Vi\~{n}als equations \cite{zv1997a} describe the dynamics of 
small amplitude Faraday waves on a deep, nearly inviscid fluid layer.  
We use the same scaling of the equations as in \cite{ss1999}, writing 
them in the form
\begin{subequations}
\label{eq:zv}
\begin{eqnarray} (\partial_{\tau}-\gamma \nabla^2)h -\Dhat\Phi & =
& \mathcal{F}(h,\Phi), \label{eq:origzv1} \\
(\partial_{\tau}-\gamma \nabla^2)\Phi -\left(\Gamma_0\nabla^2 -
G(\tau)\right)h & = & \mathcal{G}(h,\Phi), \label{eq:origzv2}
\end{eqnarray}
\end{subequations}
where $G(\tau)=G_0-f(\tau)$ and the nonlinear terms are given by
\begin{subequations}
\begin{eqnarray}
\mathcal{F}(h,\Phi) & = & -\nabla \cdot(h\nabla \Phi) +
\frac{1}{2} \nabla^2 (h^2 \Dhat \Phi) \\
& & \mbox{}- \Dhat (h \Dhat \Phi) \nonumber \\
& & \mbox{} +\Dhat \left\{h\Dhat(h\Dhat\Phi)
+\frac{1}{2}h^2\nabla^2{\Phi}\right\}, \nonumber \\
\mathcal{G}(h,\Phi) & = & \frac{1}{2}(\widehat D\Phi)^2
-\frac{1}{2}(\nabla \Phi)^2 \\
& & \mbox{} -(\widehat{ \mathcal{D}}
\Phi)\left\{h\nabla^2\Phi+\Dhat(h\widehat{ \mathcal{D}}\Phi)\right\} \nonumber \\
& & \mbox{}-\frac{1}{2}\Gamma_0\nabla\cdot\left\{(\nabla h)(\nabla
h)^2\right\}. \nonumber
\end{eqnarray}
\end{subequations}
Here $h(\vec{x},t)$ is the fluid surface height, $\Phi(\vec{x},t)$ is the 
surface velocity potential, and $\vec{x}$ is the two-dimensional spatial 
coordinate. The operator $\Dhat$ multiplies each Fourier component of a 
field by the modulus of its wave number, \emph{i.e.}, 
$\Dhat \exp{i\vec{k} \cdot \vec{x}}=|\vec{k}|\exp{i\vec{k} \cdot \vec{x}}$.

The equations depend on three dimensionless fluid parameters: the damping 
parameter $\gamma$, the gravity number $G_0$, and the capillarity 
number $\Gamma_0$.  These fluid parameters, and the dimensionless forcing 
amplitudes $f_u$ in (\ref{eq:f(t)}) are related to the physical parameters by
\begin{equation}
\label{eq:params}
\gamma \equiv \frac{2 \nu \widetilde{k}^2}{\omega}, \quad G_0 \equiv 
\frac{g_0 \widetilde{k}}{\omega^2}, \quad \Gamma_0  
\equiv \frac{\sigma \widetilde{k}^3}{\rho \omega^2}, \quad f_u\equiv  
\frac{g_u \widetilde{k}}{\omega^2}.
\end{equation}
Here $\nu$ is the kinematic viscosity, $\sigma$ is the surface  
tension, $\rho$ is the density, and $\omega$ and the $g_u$ are  
the Fourier amplitudes in the original (dimensioned) forcing function
\begin{equation}
g(t) = \sum_{u \in \mathbb{Z}^+} g_u \exp{i u \omega t}+c.c.,\quad g_u  
\in \mathbb{C}.
\end{equation}
Additionally, $\widetilde{k}$ satisfies the inviscid gravity-capillary 
wave dispersion relation
\begin{equation}
\label{eq:wavescale}
g_0 \widetilde{k}+\frac{\sigma \widetilde{k}^3}{\rho}
=\Bigl(\frac{m\omega}{2}\Bigr)^2,
\end{equation}
and $g_0$ is the usual gravitational acceleration. Note that $G_0$ 
and $\Gamma_0$ are not independent parameters since (\ref{eq:params}) 
and (\ref{eq:wavescale}) imply that
\begin{equation}
G_0+\Gamma_0=\frac{m^2}{4}.
\end{equation}
The dimensionless dispersion relation (\emph{cf.}~Eq.~\ref{eq:wavescale}) also gives 
the natural frequency $\Omega(k)$ of undamped, unforced waves as a function of their wave number $k$:
\begin{equation}
\label{eq:zvdisprel}
\Omega^2 = G_0 k + \Gamma_0 k^3.
\end{equation}

For small damping $\gamma$, (\ref{eq:zvdisprel}) provides an 
excellent estimate of the wave number which oscillates at a given 
frequency, even for forced waves; we make use of this fact in 
Section~\ref{sec:applications}.  Since the critical modes oscillate 
with dominant frequency $m/2$, we have $k_c \approx k(m/2) = 1$, where $k(\Omega)$ is the inverse of the dispersion relation from (\ref{eq:zvdisprel}).  
One may then choose a damped mode with frequency $\Omega$, find 
$k(\Omega)$, and then apply 
(\ref{eq:resonant-angle}) to estimate $\theta_{\rm res}$.

\subsection{Validity of symmetry-based results \label{sec:validity}}

To investigate the applicability of our results for finite values of 
$\gamma$, we focus on an example using three-frequency 
$(m,n,p)=(8,7,2)$ forcing and quantify the effect of the $\Omega=8-7=1$ 
damped mode; this corresponds to the penultimate entry in the second 
section of Table~\ref{tab:main-results}. Although this mode does not 
necessarily lead to the most significant resonance, we study it as an 
instructive example to address general questions about the validity 
of our symmetry results.  Damped modes which play a more important 
role are examined in the applications in Section~\ref{sec:applications}.

\begin{figure}[ht]
\centerline{\includegraphics{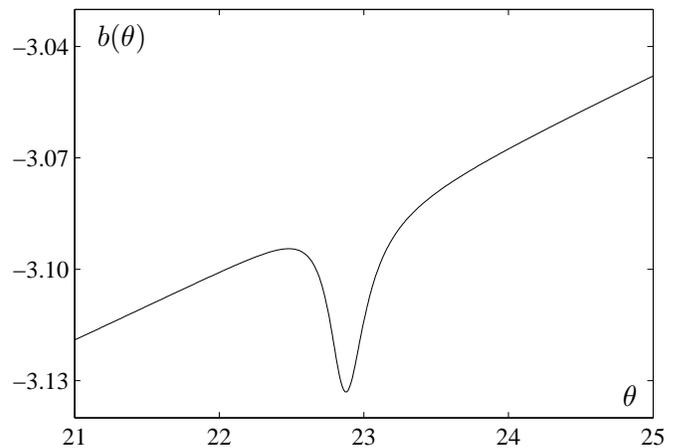}} 
\caption{Coupling coefficient $b(\theta)$ in (\ref{eq:rhombic-landau}) 
computed from (\ref{eq:zv}) using three-frequency $(m,n,p)=(8,7,2)$ forcing, 
$\gamma=0.1$, $\Gamma_0=16$, $|f_n|/|f_m|=0.4$, $|f_p|/|f_m|=0.08$, 
$\phi_m=\phi_n=\phi_p=0$.}
\label{fig:samplebtheta}
\end{figure}

From the Hamiltonian considerations in Section~\ref{sec:symmetry} 
we have $\alpha_1>0$, and thus $b_{\rm res}<0$.  We set 
$\Gamma_0=16$ in (\ref{eq:zv}), fix the ratios of the forcing 
amplitudes at $|f_n|/|f_m|=0.4$, $|f_p|/|f_m|=0.08$, and compute the 
coupling coefficient $b(\theta)$ using the method described in \cite{sts2000}.  
As predicted on the grounds of symmetry arguments, there is a dip in 
the plot of $b(\theta)$ around the angle $\theta_{\rm res} \approx 23\degree$ 
where the $\Omega = 1$ mode is in spatial resonance. An example is shown 
in Figure~\ref{fig:samplebtheta} for $\gamma=0.1$ with $\Phi$ (which 
%%%%%%%%%%%%%%%%%%%
%%%%%%%%%%%%%%%%%%%
appears in the fourth column of Table~\ref{tab:main-results}) set to $0$.

In the discussion that follows, we study various properties of $b_{\rm res}$ 
as the damping parameter $\gamma$ is varied.  In this discussion, it is important to 
realize that the results will depend on the chosen value of $m$, on which 
$\gamma$ in (\ref{eq:params}) depends indirectly through (\ref{eq:wavescale}).  When 
generalizing the results shown below to other forcing functions it is, in fact, better 
to look at the quantity $\gamma/m$ (\emph{cf.}~\ref{eq:params}).  This alternative 
nondimensional measure of the damping utilizes the critical wavenumber and the 
dominant frequency ($m\omega$, as opposed to $\omega$) and is therefore 
better suited for quantitative comparison across forcing functions with very 
different $m$ values.    We have used the scaling~(\ref{eq:params}), which 
utilizes the {\it common} frequency, to be consistent with previous 
work~\cite{ss1999,ts2002,ps2003,pts2004}.

\begin{figure}[ht]
\centerline{\includegraphics{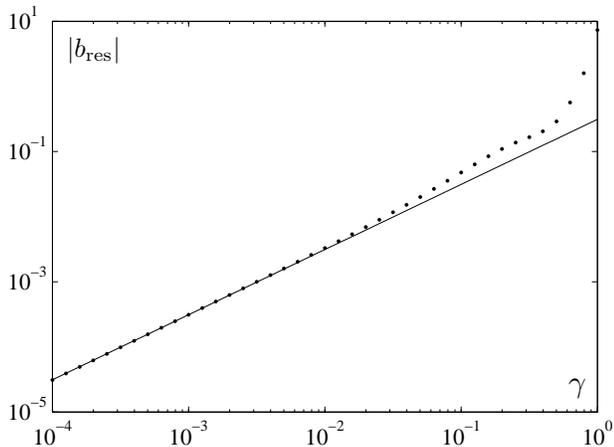}} 
\caption{Resonant contribution $b_{\rm res}$ as a function of the 
damping parameter $\gamma$.  The dots correspond to a numerical computation 
using (\ref{eq:zv}).  The straight line of slope one confirms the 
$b_{\rm res} \propto \gamma$ scaling predicted by symmetry arguments. 
The capillarity and forcing parameters used are the same as those in 
Figure~\ref{fig:samplebtheta}.}  
\label{fig:bresvsgamma}
\end{figure}

We first consider the scaling of $|b_{\rm res}|$ as $\gamma$ is varied 
with $\Phi=0$.  It follows from the result in Table~\ref{tab:main-results} that
\begin{equation}
\label{eq:scaling-one}
b_{\rm res} \propto |f_n|^2 \frac{|L_3|}{|L_3|^2-|\mu_i^2f_p^2|}.
\end{equation}
Furthermore, recall from (\ref{eq:coeff-expansions}) that 
$|L_{1,3}| \propto \gamma$ and $|L_2| \propto f_m$. Since, at the onset of 
SW, $|L_1|=|L_2|$ (see Section~\ref{sec:symmetry}), we have 
$|f_m| \propto \gamma$. Since $|f_m|$, $|f_n|$, and $|f_p|$ are held in 
a constant ratio, we also have $|f_n|, |f_p| \propto \gamma$.  Thus 
(\ref{eq:scaling-one}) becomes simply $b_{\rm res} \propto \gamma$.  
This scaling is confirmed by the numerical results of 
Figure~\ref{fig:bresvsgamma} where we hold $\Phi=0$ and plot $|b_{\rm res}|$
%%%%%%%%%%%%%%%%%%%%%%%%%%
%%%%%%%%%%%%%%%%%%%%%%%%%% 
as a function of $\gamma$.  The numerical data is shown as points.  For 
comparison, a line of slope one is drawn through the first data point, 
confirming the proportionality to $\gamma$.  The theoretically predicted 
scaling holds reasonably well up to $\gamma \sim \order(10^{-1})$, and the 
numerical result does not strongly diverge from the prediction 
until~$\gamma \approx 0.5$.

\begin{figure}[ht]
\centerline{\includegraphics{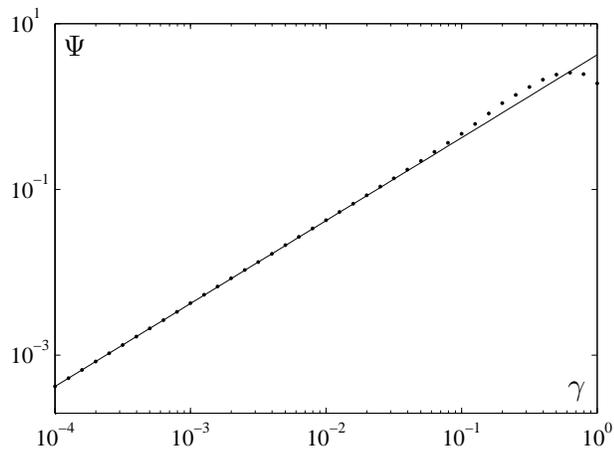}}
\caption{Half-width $\Psi$ of the resonant ``dip'' as a function of the 
damping $\gamma$. The dots correspond to a numerical computation using 
(\ref{eq:zv}).  The straight line of slope one confirms the predicted 
$\Psi \propto \gamma$ scaling.  The capillarity and forcing parameters 
used are the same as those in Figure~\ref{fig:samplebtheta}.}
\label{fig:psivsgamma}
\end{figure}

Next, we examine the scaling of the half-width $\Psi$ of the dip 
at $\theta=\theta_{\rm res}$.  For $\theta \neq \theta_{res}$, the natural 
frequency of $k_d$ will differ from the resonant frequency ($m$, 
$2m$, $n$, etc.).  At leading order, this {\it detuning} appears in the 
coefficient $L_3$ as an imaginary part, \emph{i.e.}, 
$L_3 = -\varrho_r \gamma + i\varrho_i$ 
(\emph{cf.}~Eq. \ref{eq:coeff-expansions}c).  If the detuning is 
small, the linear approximations $\varrho_i \propto k_d - k_{\rm res} 
\propto \theta - \theta_{\rm res}$ can be used (here $k_{\rm res}$ is the 
wave number associated with the resonant frequency) and so 
$\varrho_i = c(\theta-\theta_{\rm res})$ for some real constant $c$, 
\emph{i.e.},
\begin{equation}
L_3 \approx -\varrho_r\gamma + ic(\theta-\theta_{\rm res}).
\end{equation}
Substituting this expression into the result from Table~\ref{tab:main-results} 
shows that $\Psi \propto \gamma$. Numerical results are displayed as points 
on the log-log plot in Figure~\ref{fig:psivsgamma}.  For comparison, we 
%%%%%%%%%%%%%%%%%%%%%%%
%%%%%%%%%%%%%%%%%%%%%%%
plot a line of slope one fit through the first data point.  As with the 
dip magnitude $|b_{\rm res}|$, the theoretical prediction remains reasonable 
up to $\gamma \sim \order(10^{-1})$.

\begin{figure}
\centerline{\includegraphics{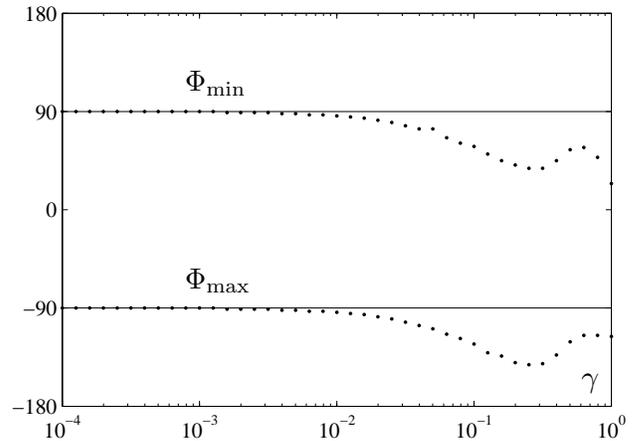}} 
\caption{The values of $\Phi$ at which $b(\theta_{\rm res})$ takes on its 
minimum and maximum values as a function of the damping $\gamma$.  The dots 
correspond to numerical data, while the lines at $90\degree$ and $-90\degree$ show the 
predicted minimum and maximum respectively. The capillarity and forcing amplitudes used 
are the same as those in Figure~\ref{fig:samplebtheta}.}  
\label{fig:phiextremavsgamma}
\end{figure}

\begin{figure}
\centerline{\resizebox{\columnwidth}{!}{\includegraphics{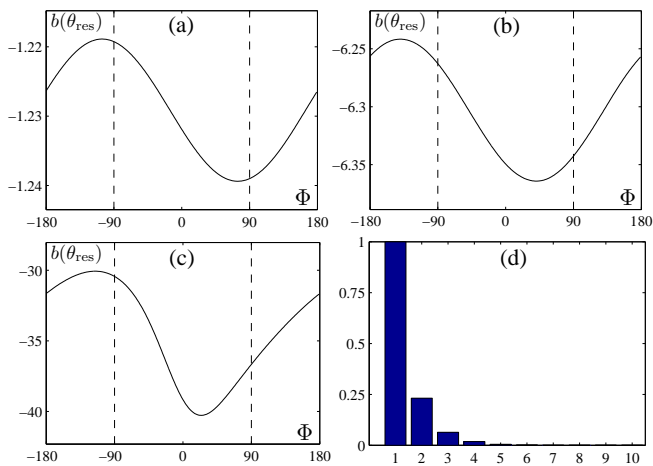}}} 
\caption{(a -- c) Dependence of $b(\theta_{\rm res})$ on the phase $\Phi$. 
(a) With damping $\gamma=0.04$. As predicted by the symmetry arguments in 
Section~\ref{sec:symmetry}, the phase dependence is sinusoidal with 
minimum and maximum near $\pm 90\degree$. (b) $\gamma=0.2$. The 
phase dependence is sinusoidal, but there is a phase shift of 
approximately $45 \degree$. (c) $\gamma=0.1$. The dependence is 
no longer sinusoidal. (d) Fourier transform of the data in (c). The zero 
component has been removed and the remaining data has been normalized so 
that the strongest component has magnitude one. The dependence on higher 
harmonics, \emph{e.g.} $2\Phi,3\Phi,4\Phi$ is apparent. For all plots, the 
capillarity and forcing amplitudes are as those in Figure~\ref{fig:samplebtheta}.}  
\label{fig:sinusoidal}
\end{figure}

Finally, we consider the dependence of $b(\theta_{\rm res})$ on 
$\Phi=\phi_2-2\phi_8+2\phi_7$, and examine how this $\Phi$ dependence 
changes with increasing $\gamma$.  From Table~\ref{tab:main-results}, we 
expect that the dependence is sinusoidal and, from the fact that $\mu_i >0$ 
\cite{ps2003} for (\ref{eq:zv}), we anticipate $b(\theta_{\rm res})$ reaching a 
maximum (\emph{i.e.}, having the shallowest dip) near 
$\Phi = \Phi_{\rm max} = -90\degree$ and reaching a minimum (\emph{i.e.}, 
having the deepest dip) near $\Phi = \Phi_{\rm min} = 90\degree$.  
Figure~\ref{fig:phiextremavsgamma} shows how the numerically calculated 
values (dots) of $\Phi_{\rm min}$ and $\Phi_{\rm max}$ differ from the theoretical 
predictions (lines) as $\gamma$ is increased.
%%%%%%%%%%%%%%%%%%%%%%%%%%
%%%%%%%%%%%%%%%%%%%%%%%%%
To elucidate the departure from the theoretical prediction, we show three 
profiles corresponding to three different values of $\gamma$ in 
Figure~\ref{fig:sinusoidal}. In Figure~\ref{fig:sinusoidal}a, $\gamma=0.04$
%%%%%%%%%%%%%%%%%%%%%%%%%
%%%%%%%%%%%%%%%%%%%%%%%% 
and the profile, as predicted, appears sinusoidal. The vertical lines drawn at 
$\pm 90 \degree$ agree reasonably well with the numerically calculated maximum 
and minimum.  In Figure~\ref{fig:sinusoidal}b, $\gamma=0.2$ and, although 
the profile is still sinusoidal, it is shifted by approximately $45\degree$ 
with respect to the theoretical prediction.  In Figure~\ref{fig:sinusoidal}c, 
$\gamma=1$, and the profile no longer resembles a sine function. This is 
demonstrated further by the plot in Figure~\ref{fig:sinusoidal}d, which 
shows the Fourier transform of the data in Figure~\ref{fig:sinusoidal}c.  
The zero component (\emph{i.e.}, the $\Phi$-independent part) has been 
removed, and the remaining data has been normalized so that the strongest 
component has magnitude one. The data indicate that higher harmonics 
of $\Phi$ are now important.  Note that the phase shift appears well before 
the higher harmonics come into play (see Figure~\ref{fig:sinusoidal}b), a 
fact that can be understood as follows.  The $\Phi$ dependence in  
Table~\ref{tab:main-results} originates with the phase of 
terms in the normal form reduction, and depends on products of 
the coefficients in~(\ref{eq:coeff-expansions}).  If the next order 
terms in the expansions describing these coefficients are kept, a phase 
shift of $\order(\gamma)$ is obtained.  In contrast, higher harmonics of 
$\Phi$ are generally associated with higher order (as opposed to next 
order) terms in the expansions~(\ref{eq:coeff-expansions}).  This is a 
result of time-translation symmetry, which requires that terms involving 
additional powers of the forcing amplitudes $f_u$ only appear in 
certain combinations.  The specific order in $\gamma$ at which these 
new terms become relevant depends in complicated fashion on 
the particular choice of forcing frequencies.

In this section we have explored the validity of our symmetry results 
with respect to the small $\gamma$ assumption under which they were 
derived.  For small $\gamma$, the symmetry results are in excellent 
agreement with the numerical ones.  For larger $\gamma$ , the scalings 
predicted by symmetry are not correct. However, many of the important 
\emph{qualitative} features are preserved.  In particular, even at larger 
$\gamma$, increasing $\gamma$ increases $|b_{\rm res}|$.  Furthermore, 
even though the dependence of $b_{\rm res}$ on $\Phi$ is no longer 
sinusoidal, there are still special phases $\Phi_{\rm min}$ and $\Phi_{\rm max}$ 
which minimize and maximize $b_{\rm res}$, suggesting that even in experiments 
with large damping, tuning the forcing phases may be an effective means 
by which to control resonant triad interactions important to pattern formation.

\section{Applications \label{sec:applications}}

The results in Table~\ref{tab:main-results} may be used to understand -- 
and control -- certain phenomena in Faraday systems.  For each of the 
following examples, we apply our symmetry-based methods and demonstrate 
the results via numerical calculations using (\ref{eq:zv}).

\subsection{$\ratio{1}{2}$ temporal resonance and impulsively forced  
{F}araday waves\label{sec:delta}}

We focus on the cases for which $\Omega=m$ in Table~\ref{tab:main-results}, 
so that the the critical modes and the damped mode are in a 
$1\colon\negthinspace2$ temporal resonance. From the Hamiltonian 
considerations in Section~\ref{sec:symmetry}, $\alpha_1>0$ and thus 
$b_{\rm res}<0$.  Also, recall from Section~\ref{sec:discussion} that for 
this case, the modes are coupled at $\order(1)$. Therefore, the 
contribution $b_{\rm res}$ is $\order(\gamma^{-1})$, which is larger 
than for the other cases, where $b_{\rm res}$ is only 
$\order(\gamma)$. In short, the $\Omega=m$ mode has a very strong 
influence on $b(\theta)$. The implications of this well-known resonance 
for Faraday waves have been investigated in a number of studies, 
including \cite{zv1997a}.

When $f_{2m}$ forcing is present, the size of $b_{\rm res}$ depends 
on the phase $\Phi = \phi_{2m}-2\phi_m$; see the first entry in the 
second section of Table~\ref{tab:main-results}. This phase dependence 
has previously been calculated in \cite{zv1997b} by means of a 
perturbation expansion on the Zhang-Vi\~{n}als model (\ref{eq:zv}).  
Our work confirms the phase dependence in a model-independent manner, 
exclusively by means of symmetry considerations. The phase dependence 
gives us a convenient and powerful means by which to control the 1:2 
resonance and influence the shape of $b(\theta)$. In particular, 
using $\Phi=90^\circ$ maximizes the effect of the resonance, while 
$\Phi=-90^\circ$ minimizes it.

\begin{figure*}                                                                                                                                              
\centerline{\includegraphics{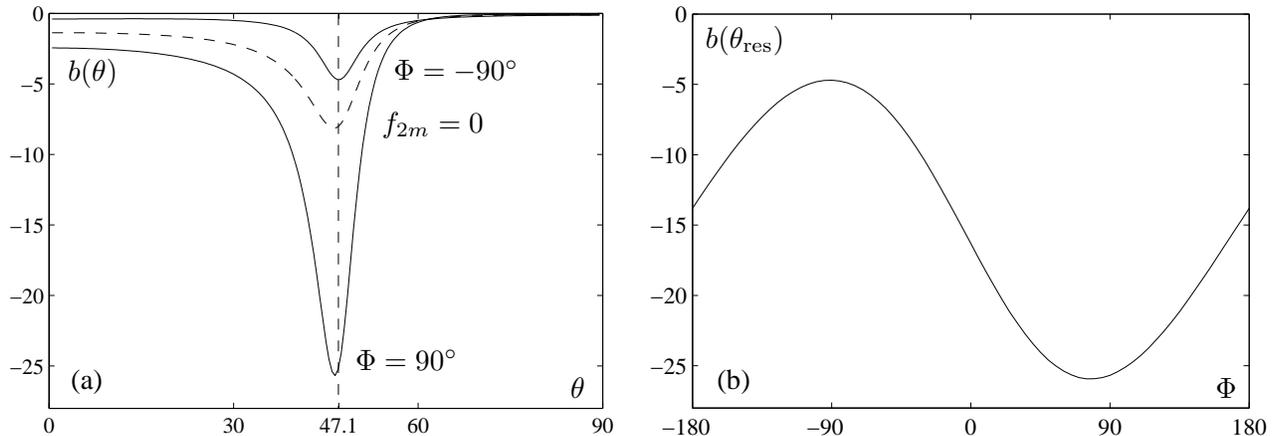}}
\caption{\label{fig:12} Effect of relative forcing phase on the first  
harmonic resonance, \emph{i.e.}, resonance with the $\Omega = m$ mode,  
for $(m,n)=(1,2)$ forcing. The relevant phase $\Phi$ is given in Table  
\ref{tab:main-results}. (a) Cross-coupling coefficient $b(\theta)$ with  
$\Phi=90^\circ$ and $\Phi=-90^\circ$; the single frequency case  
(dashed line) is shown for reference. (b) Dip magnitude 
$b(\theta_{\rm res})$ versus $\Phi$.  For these calculations, the 
parameters in (\ref{eq:zv}) are $\gamma = 0.008$ and $\Gamma_0=0.125$, 
and the forcing amplitude ratio is $|f_n|/|f_m|=0.396$.}
\end{figure*}

In Figure~\ref{fig:12} we show a numerical example for $(m,n)=(1,2)$ 
%%%%%%%%%%%%%%%%%%%%%%%%%%%%%
%%%%%%%%%%%%%%%%%%%%%%%%%%%%%
forcing.  The parameters in (\ref{eq:zv}) are $\gamma = 0.008$
and $\Gamma_0=0.125$. The forcing amplitude ratio is $|f_n|/|f_m|=0.396$, 
which is far from the codimension-two point $|f_n|/|f_m|=3.53$ at which 
waves with dominant frequency $n/2$ set in.
The $\Omega=m$ mode has wave number $k(m) \approx 1.83$, and thus $\theta_{\rm res}
\approx 47.1\degree$. Consistent with Table~\ref{tab:main-results}, a dip
in $b(\theta)$ is found at this angle. As predicted, by choosing
$\Phi=90\degree$, we achieve the largest dip at $\theta_{\rm res}$ and
thus a strong suppression of patterns involving angles near this one.
On the other hand, using $\Phi$ near $-90^\circ$, actually reduces the
effect of the triad interaction by a factor 
$1/2 <|L_3|/(|L_3|+|\tilde{\mu}_i f_{2\Omega}|) < 1$ relative to
the single-frequency case, so the suppression is much weaker.

As discussed in Section~\ref{sec:background}, the spatiotemporal 
resonances we consider in this paper may also affect the self-interaction 
coefficient $a$ in the one-dimensional analogue of (\ref{eq:rhombic-landau}), 
namely (\ref{eq:1dsw}). 
In the case of the $\ratio{1}{2}$ temporal resonance, the condition  
$\Omega(k_d)=2\Omega{k_c}$ must be satisfied along with (\ref{eq:resonance-1d}).  
There will then be a contribution to the self-interaction coefficient $a$ 
in (\ref{eq:1dsw}) whose dependence on the forcing and damping parameters 
is precisely that given in Table~\ref{tab:main-results}.  In practice, one may  
vary the frequency $\Omega$ by tuning the capillarity number $\Gamma_0$ 
which appears in the dispersion relation~(\ref{eq:zvdisprel}).  In an 
experiment, this might be achieved by varying the base forcing frequency 
$\omega$ (see  Eq.~\ref{eq:params}). 

The results of Table~\ref{tab:main-results} for the $\ratio{1}{2}$ 
spatiotemporal resonance and its effects on the self-interaction coefficient 
$a$ may be used to understand certain features of impulsively-forced Faraday 
waves, \emph{i.e.}, waves forced by a periodic sequence of impulses rather 
than a smooth forcing function of the form of (\ref{eq:f(t)}).  Impulsive 
forcing was studied first in \cite{bj1996} and subsequently in \cite{cps2004}. 
The original motivation for studying impulsive forcing is that it allows for a purely 
analytic linear stability calculation for arbitrarily large damping, 
in contrast to the case of smooth periodic forcing functions for which the 
linear analysis must be performed numerically \cite{kt1994,bet1996} and analytic 
approximations are only valid for small damping.

\begin{figure}
\centerline{\includegraphics{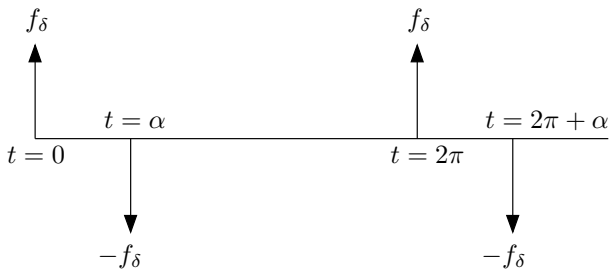}}
\caption{\label{fig:deltaforce} Schematic representation of the asymmetric  
$\delta$-function forcing specified by (\ref{eq:deltaforce}).}
\end{figure}

In \cite{cps2004} the forcing function takes the form
\begin{equation}
\label{eq:deltaforce}
f(t) = f_\delta \sum_{n=0}^{\infty} \delta(t-2\pi n) - \delta(t-2\pi n  
- \alpha),
\end{equation}
representing an alternating sequence of $\delta$-functions of strength
$f_\delta$. The sequence has a temporal asymmetry controlled by the
parameter $\alpha \in (0,2\pi)$, which determines the amount of time 
between a positive pulse and the subsequent negative pulse.  A 
depiction of (\ref{eq:deltaforce}) is shown in Figure~\ref{fig:deltaforce}.  
%%%%%%%%%%%%%%%%
%%%%%%%%%%%%%%%%
In \cite{cps2004}, $a(\Gamma_0)$ is calculated from (\ref{eq:zv}), and a 
large dip at $\Gamma_0=\Gamma_{\rm res}$ is observed, where 
$\Gamma_{\rm res}$ is the parameter value for which the $\ratio{1}{2}$ 
spatiotemporal resonance is satisfied. For $\gamma$ sufficiently small, 
it is noted that this dip becomes more negative (\emph{i.e.}, the 
corresponding $a_{\rm res}$ becomes more negative) as the asymmetry 
parameter $\alpha$ is varied across the interval $(0,2\pi)$. This observation 
is consistent with the results in Table~\ref{tab:main-results}, as we now explain.

From Table~\ref{tab:main-results}, there are at most two forcing 
frequencies which affect the $\Omega=m$ damped mode at leading order, 
namely $m$ and $2m$.  We therefore consider a drastic truncation of the 
Fourier series for the forcing function~(\ref{eq:deltaforce}), keeping  
the first two terms, which are the only terms affecting the resonance 
at leading order:
\begin{equation}
\label{eq:deltatrunc}
f(t) = f_1 \exp{it} + f_2 \exp{2it}+c.c. \,,
\end{equation}
where
\begin{equation}
f_1 = \frac{f_\delta}{2\pi} \left(1-\exp{-i\alpha}\right), \quad 
\frac{f_\delta}{2\pi}\left(1-\exp{-2i\alpha}\right). 
\end{equation}
From (\ref{eq:deltatrunc}), we see that varying $\alpha$ affects both the 
amplitudes and the phases of the forcing components.  For (\ref{eq:zv}) with 
weak damping and forcing, and for the two-frequency 
truncation~(\ref{eq:deltatrunc}), the Faraday instability occurs when 
$|f_1|=\gamma$ (this follows directly from the results in \cite{ts2002}). 
By setting $f_\delta$ equal to its critical value and making a translation 
in time, we can write the forcing function at onset as
\begin{equation}
f^{crit}(t) = \gamma \exp{it} + F_2 \exp{2it}+c.c. \,,
\end{equation}
where
\begin{equation}
\label{eq:F2}
F_2 = -2i\gamma\cos\left( \frac{\alpha}{2}\right).
\end{equation}
The first entry in the second section of Table~\ref{tab:main-results} indicates   
that the $\ratio{1}{2}$ spatiotemporal resonance produces a negative contribution 
$a_{\rm res}$ to the self-interaction coefficient given by $-\alpha_1 P_2(\Phi)$ 
where $\Phi=\arg(F_2)$.  Using the expression (\ref{eq:F2}) and 
simplifying reveals that
\begin{equation} \label{eq:deltaresult}
a_{\rm res} = -\frac{\alpha_1}{|L_3|+ 2 |\mu_i| \gamma \cos(\alpha/2)},
\end{equation}
which decreases as 
$\alpha$ is increased across the interval $(0,2\pi)$. This is consistent with the observation in \cite{cps2004}, which successfully fits numerical results to this functional form, at least for small $\gamma$.  From (\ref{eq:1dsw}) 
we see that the periodic striped state has a steady state amplitude of 
$|A_1|^2 = -\lambda/a$. Thus, experimentally, the wave height may be 
controlled by varying $\alpha$. Larger $\alpha$ causes smaller $a$ and, 
consequently, larger amplitude waves.

\subsection{Stabilization of superlattice patterns with 
multi-frequency forcing\label{sec:sl}}

We now generalize the simple one-dimensional example just presented.  Our 
symmetry-based results suggest a methodology for ``engineering'' specific 
two-dimensional patterns through a judicious choice of forcing function.  The 
idea is to exploit the results in Table~\ref{tab:main-results} in constructing 
a multi-frequency forcing function such that enhancing (and/or suppressing) 
resonances occur at carefully chosen angles.  We will apply this methodology 
to demonstrate how a superlattice pattern of the SL-I type observed 
in~\cite{kpg1998} may be stabilized. Stabilization of this superlattice 
patterns can be related to the damped $\Omega = n-m$ 
``difference frequency'' mode in Table~ \ref{tab:main-results}.  A 
demonstration is provided in~\cite{sts2000}, and further explorations are 
performed in \cite{ts2002,pts2004}.  The method we outline below, however, 
results in a dramatically more pronounced stabilization than was obtained in 
previous work. In particular, it can lead to stable superlattice patterns at onset of the primary instability of the flat fluid surface.

\begin{figure}[ht]
\centerline{\includegraphics{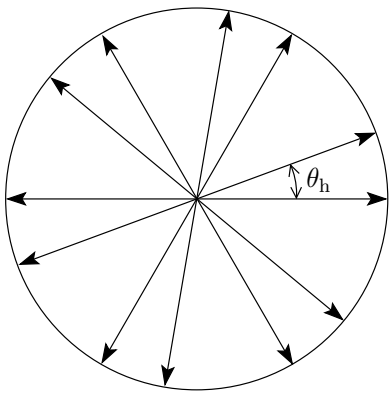}}
\caption{\label{fig:slmodes} Schematic of the Fourier wave vectors 
corresponding to the 12 dominant waves which comprise an SL-I superlattice 
pattern. The vectors point to the vertices of two hexagons, one rotated 
by an angle $\theta_{\rm h} < 30^\circ$ with respect to the other.}
\end{figure}

\emph{Step 1:} {\it Use geometry to determine the angles for the 
desired enhancing (or suppressing) effects}.  For the SL-I 
pattern, the twelve dominant waves making up this pattern have wave 
vectors that lie at the vertices of two hexagons, one rotated by an angle 
$\theta_{\rm h} < 30^\circ$ with respect to the other; see 
Figure~\ref{fig:slmodes}. The stability of this pattern may be studied 
%%%%%%%%%%%%%%%%%%%
%%%%%%%%%%%%%%%%%%%
within the framework of a twelve-dimensional bifurcation problem which 
describes the competition of these superlattice patterns with stripes, 
rhombic patterns, and hexagons. This approach is developed in 
\cite{dg1992,dss1997,sp1998} (the full bifurcation equations may be 
found there).  A key result is that the stability of the superlattice 
pattern associated with $\theta_h$ depends on coefficients in the 
bifurcation equations which we call $(b_4,b_5,b_6)$, where
\begin{subequations}
\label{eq:bcoefs}
\begin{eqnarray}
b_4 & = & b(\theta_h)/|a|, \\
b_5 & = & b(60\degree - \theta_h)/|a|, \\
b_6 & = & b(60\degree + \theta_h)/|a|,
\end{eqnarray}
\end{subequations}
with $a$ and $b(\theta)$ appearing in (\ref{eq:rhombic-landau}). In 
particular, the superlattice pattern is favored when $(|b_4|,|b_5|,|b_6|)$ 
are all small. Since $b(\theta)$ may be made small in magnitude with 
``enhancing'' resonances that cause spikes in $b(\theta)$, geometry 
dictates that we should arrange for such resonances to occur at one or 
more of the angles $\theta$, $60\degree - \theta$, $60\degree + \theta$. 
For a more detailed discussion, see \cite{sts2000}.

\emph{Step 2:} {\it Use the dispersion relation and appropriate resonance 
conditions from Table~\ref{tab:main-results} to find a good set of forcing 
frequencies which satisfy the geometrical constraints from Step 1.}  For our 
SL-I example, since we want to construct enhancing resonances, we turn 
our attention to the $\Omega = n-m$ ``difference frequency'' mode.   We begin with 
three-frequency $(m,n,p)$ forcing, aiming to make two of $(|b_4|,|b_5|,|b_6|)$ 
small using the two difference frequency modes $\Omega=n-m$ and $\Omega=p-m$.  
We choose to stabilize a superlattice pattern having $\theta_h \simeq 20.3 \degree$ 
(this is a different SL-I pattern than the one observed in \cite{kpg1998}, but it is 
in the same family of patterns; see \cite{dss1997,sp1998}).   The two wave numbers corresponding to 
the difference frequency modes satisfy the resonance conditions
\begin{subequations}
\label{eq:diffks}
\begin{eqnarray}
\Omega^2(k_{n-m}) & = & (n-m)^2, \\
\Omega^2(k_{p-m}) & = & (p-m)^2.
\end{eqnarray}
\end{subequations}
With the optimal wave numbers for these damped modes dictated by geometry, the 
aim is to find a set of forcing frequencies $(m,n,p)$ such that $k_{n-m}$ and 
$k_{p-m}$ of (\ref{eq:diffks}) are as close to the optimal wave numbers 
as possible.   In practice we also vary $\Gamma_0$ so as to arrange for frequencies 
$(m,n,p)$ that are not too large - this is not strictly necessary but it eases our 
numerical computations to use smaller sets of integers.   In this case we obtain 
reasonable agreement by using $(m,n,p)=(8,10,11)$ and $\Gamma_0=5.24$.  
The wave numbers predicted by (\ref{eq:diffks}) are 
$(k_{n-m},k_{p-m}) \simeq (0.351,0.682)$ and the corresponding resonance 
angles of (\ref{eq:resonant-angle}) are $(\theta_{n-m},\theta_{p-m}) 
\simeq (159.8\degree,140.1\degree)$.   These will cause spikes in $b(\theta)$ at 
appropriately $20.2\degree$ and $39.9\degree$, respectively; note that the latter 
angle is close to $60\degree - 20.3 \degree$ (\emph{cf.}~Eq.~\ref{eq:bcoefs}b). 

\begin{figure}[ht]
\centerline{\includegraphics{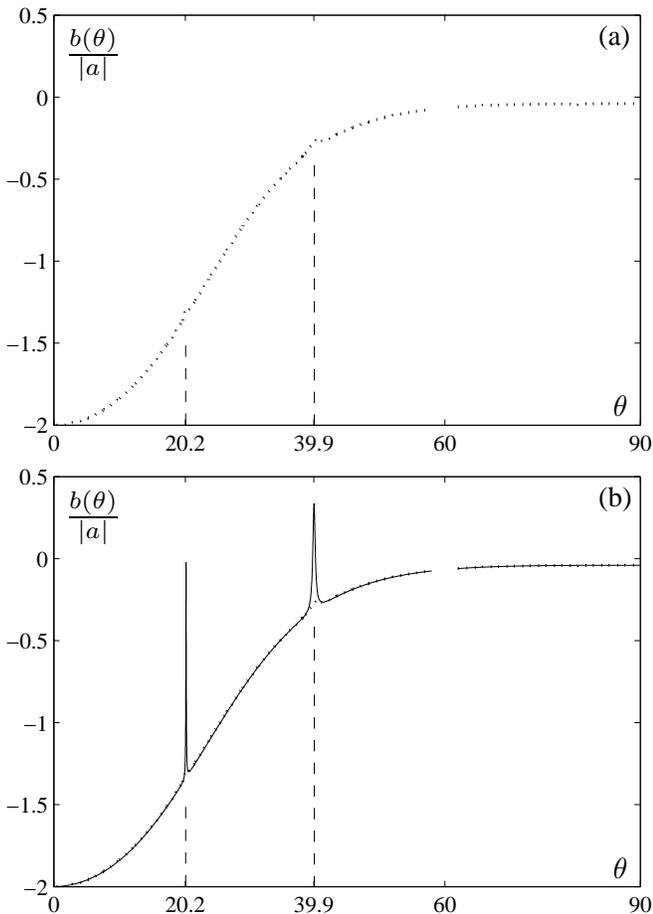}} 
\caption{(a) Coupling coefficient for computing \mbox{superlattice-I} pattern 
stability. We use three-frequency forcing with 
$(m,n,p)=(8,10,11)$. The  two small ``bumps'' at 
$(\theta_{n-m},\theta_{p-m})=(20.3\degree,39.9\degree)$ 
are due to resonance with the modes oscillating with the difference 
frequencies $\Omega = n-m$ and $\Omega=p-m$.  No superlattice patterns are 
stable.  (b) Like (a), but with additional forcing  
frequency components $(q,r)=(4,6)$ which parametrically force the  
difference frequency modes. The result from (a) is duplicated as a dotted line for comparison. The two bumps become two very large spikes,  
and the superlattice pattern with angle $\theta_h \simeq 20.3\degree$  
is stabilized.  The fluid parameters used are $\gamma = 0.1$ and 
$\Gamma_0 = 5.24$. The forcing amplitude ratios and phases used are 
given in the text.  The region around $60 \degree$ corresponds to a 
hexagonal interaction not captured by our calculation, and thus has 
been removed.} 
\label{fig:slstab}
\end{figure}

We compute the coupling coefficient from (\ref{eq:zv}) with damping 
$\gamma=0.1$, forcing amplitude ratios $|f_n|/|f_m| = 1.54$, $|f_p|/|f_m| = 1.85$,
and forcing phases $(\phi_m,\phi_n,\phi_p)=(0\degree,0\degree,0\degree)$.   
The forcing ratios were chosen to make $b_{\rm res}$, which is proportional 
to $|f_n|^2$ in one case and $|f_p|^2$ in the other, as large as possible, while 
at the same time avoiding the critical values (\emph{i.e.}, the modes oscillating 
at $n/2$ and $p/2$ must remain damped).  The coefficients $(b_4,b_5,b_6)$ are 
represented in Figure~\ref{fig:slstab}a,
%%%%%%%%%%%%%%%%%%%%
%%%%%%%%%%%%%%%%%%% 
where we plot $b(\theta)/|a|$ as a dotted line.  As expected, there are two  
bumps due to the two difference frequency resonances, though they are  
quite small (the large dip around $\theta=0\degree$ is due to resonance 
with the $\Omega=m$ mode). In fact, though the observed resonances at  
$20.3\degree$ and $39.9\degree$ are in excellent agreement with the 
prediction, the effect is far too weak to stabilize a pattern at the chosen angle, and so more work must be done.

\emph{Step 3:} {\it Use the results in Table~\ref{tab:main-results}  
to further enhance/suppress the nonlinear interactions.}  In this case we add 
the forcing components $(q,r)=(4,6)$ in order to parametrically force the 
damped $\Omega=2$ and $\Omega=3$ difference frequency modes and 
obtain larger $|b_{\rm res}|$.  In order to favor our chosen SL-I pattern, we choose 
$|f_n|/|f_m|$ and $|f_p|/|f_m|$ as before, and take $|f_q|/|f_m| = 0.184$ and 
$|f_r|/|f_m| = 0.505$. These ratios are close to (but below) their critical values when $|f_m|=|f_m|^{\rm crit}$. We have chosen the phases to be $(\phi_m,\phi_n,\phi_p,\phi_q,\phi_r)=(0\degree,0\degree,0\degree,
-7\degree,-10\degree)$. Though the arguments of Section \ref{sec:background} suggest that we should make $b_{\rm res}$ as large and positive as possible to favor the pattern, we are working with a cubic truncation of the bifurcation equations and so we actually want $b_{\rm res}$ such that  $|b|$ is very small (as previously stated).  We might have adjusted the forcing amplitude ratios to achieve this situation, but instead, we find it more convenient to vary the forcing phases away from the optimal values predicted by Table \ref{tab:main-results}.

The coupling coefficient appears as the solid line
in Figure~\ref{fig:slstab}b. It nearly duplicates the result from the  
three-frequency case (which is included as a dotted line for comparison)
but the two small bumps have become large spikes.  
We find that at $\theta=20.3\degree$, $(b_4,b_5,b_6)=(0.02230,-0.01887,-0.0045)$. 
To study the stability of the superlattice states, we perform a bifurcation 
analysis using the overall forcing strength $f_{tot}\equiv\sqrt{|f_m|^2+|f_n|^2 
+ |f_p|^2 + |f_q|^2 + |f_r|^2}$ as the bifurcation parameter.  A branch of superlattice patterns with 
$\theta_h \simeq 20.3\degree$ bifurcates transcritically from the trivial state, 
and the subcritical branch then turns around in a saddle-node bifurcation 
at a particular value $f_{tot} = f_{SN}$.  At a slightly greater forcing strength $f_{SL} > f_{SN}$ (still in the subcritical regime), the superlattice pattern is stabilized, and remains stable for $f_{tot}>f_{SL}$ (at least within the realm of validity of the weakly 
nonlinear description provided by the bifurcation equations).

The methodology here is more successful than our previous attempts at stabilizing 
superlattice patterns.  Our work in \cite{sts2000} created a spike at only 
one angle (as opposed to two, as here) and that in \cite{ts2002} 
did not parametrically force the damped mode and did not make an appropriate 
choice of phases in order to optimize the size of the spike. By combining 
multiple resonances with appropriately chosen phases, we have used 
Table~\ref{tab:main-results} to obtain dramatically increased stabilization 
of the desired pattern.

\subsection{A conjecture on quasipatterns \label{sec:quasi}}

The superlattice pattern discussed above belongs to one intriguing class 
of complex patterns; another such class is that of quasipatterns.  
Quasipatterns are the continuum analogues of quasicrystals.  Unlike the 
superlattice patterns, they are not spatially periodic.  However, their 
Fourier spectra possess discrete rotational symmetry.  Quasipatterns 
have been observed in a number of Faraday wave experiments, including 
\cite{cal1992,ef1994,kpg1998}.

A common approach to certain types of quasipatterns has been to describe 
them using amplitude equations for the evolution of a number of critical 
modes equally spaced around a critical circle in Fourier space; see, for example, 
\cite{np1993,zv1997a,er2001}.  Recent work in \cite{rr2003} elucidates 
the technical problems with this approach.  The issue is that through nonlinear 
interactions, the critical modes generate other modes which come arbitrarily 
close to the critical circle, and a center manifold reduction to a finite dimensional bifurcation problem is not possible.  
The usual amplitude equation description is thus without a rigorous mathematical 
foundation. Nonetheless, our basic physical ideas should still apply to quasipatterns.  
We may tune our forcing function to drive energy into modes corresponding 
to different resonant angles and thus favor the corresponding patterns.

\begin{figure}[ht]
\centerline{\includegraphics{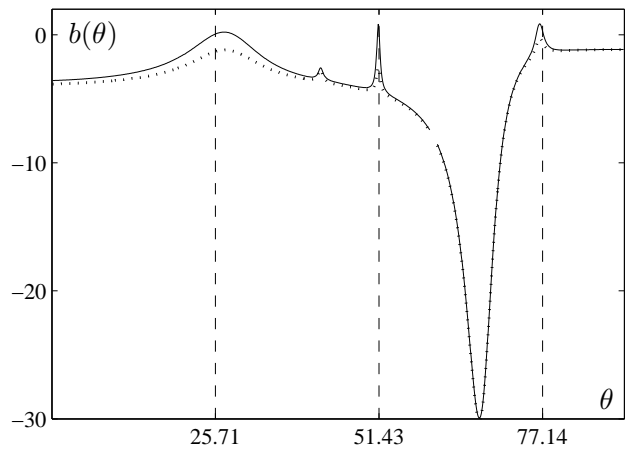}}
\caption{Coupling coefficient $b(\theta)$ in (\ref{eq:rhombic-landau}) as 
computed from (\ref{eq:zv}) for seven-frequency $(m,n,p,q,r,s,t)=
(12,17,20,27,10,16,30)$ forcing. The forcing phases are all $0$ for the 
dotted line, while the solid line corresponds to the optimal choice 
$\phi_{r,s,t}=90\degree$.  The three damped modes with frequencies $n-m$, 
$p-m$, and $q-m$ are parametrically forced by the $(r,s,t)$ components. 
These three difference frequencies lead to spikes in $b(\theta)$ at angles 
which may help stabilize a 14-fold quasipattern; the desired locations of 
these spikes (as determined by geometry) are indicated by dashed vertical lines.  
The small spike around $42\degree$ is due to another difference frequency 
resonance not of interest here.  As in Figure \ref{fig:slstab}, the region 
around $60 \degree$ has been removed.}
\label{fig:quasi}
\end{figure}

For example, here we suggest a forcing function which may favor a 14-fold 
quasipattern, which, to date, has not been observed in Faraday wave 
experiments. We use the methodology outlined in the previous example. We 
wish to arrange for $b_{\rm res} > 0$ at the angles 
$\theta_j = j (180\degree) /14$, $j=1\ldots 3$. This actually accounts 
for twelve of the angles in the pattern, since by symmetry, 
$b(\theta) = b(180\degree - \theta) = b(180\degree + \theta)$. We choose the 
seven-frequency forcing function $(m,n,p,q,r,s,t) = (12,17,20,27,10,16,30)$ with 
the capillarity parameter $\Gamma_0 = 28.8$ in (\ref{eq:zv}).  
The $\Omega = n - m = 5$, $\Omega = p - m = 8$, and $\Omega = q - m = 15$  
difference frequency modes are parametrically forced by the $(r,s,t)=(10,16,30)$ 
components. We take forcing frequency ratios $|f_n|/|f_m| = 1.2$, 
$|f_p|/|f_m| = 1.6$, $|f_q|/|f_m| = 2.8$, $|f_r|/|f_m| = 0.62$, 
$|f_s|/|f_m| = 1.2$ and $|f_t|/|f_m| = 2.2$ and compute $b(\theta)$ 
from (\ref{eq:zv}). The results are shown in Figure~\ref{fig:quasi}. 
%%%%%%%%%%%%%%%%%%%
%%%%%%%%%%%%%%%%%%%%
The dotted line corresponds to the naive choice of zero for all of 
the forcing phases. The solid line corresponds to the optimized case 
prescribed by Table~\ref{tab:main-results}, namely 
$\phi_r=\phi_s=\phi_t=90\degree$. In both cases, the three difference 
frequency modes cause spikes in $b(\theta)$ at the required angles.

\section{Conclusions \label{sec:conclusions}}

In this paper, we have used methods from equivariant bifurcation theory 
to study resonant triad interactions in Faraday waves.  We have shown how the 
spatial and weakly broken temporal symmetries (or alternatively, parameter 
symmetries) may be used to determine which spatiotemporally resonant damped 
modes play the most important roles in pattern selection.  The symmetry-based 
analysis not only identifies the modes, but tells us how the strength of the 
triad interactions depend on the frequencies, amplitudes, and relative 
phases of the various components in an arbitrary multi-frequency periodic 
forcing function.  In many cases we know whether the interaction has an enhancing 
or suppressing effect on associated patterns.  The study in this paper constitutes a 
somewhat unusual situation (we know of only a few others, such as 
\cite{c2001}) because significant information about the bifurcation 
coefficients, namely their scaling with respect to the physical parameters and in 
some cases their sign, can be obtained without resorting to calculations 
using the governing equations. This is possible because of the structure  
imposed by the parameter symmetries of the problem.

We have applied our results to impulsively forced and multi-frequency forced 
Faraday waves in several examples, emphasizing how the resonant interactions 
can be controlled by choosing judiciously the parameters in the forcing 
function $f(\tau)$. An appropriate choice allows one to stabilize complex 
patterns such as the superlattice-I pattern examined in 
Section~\ref{sec:applications}.  Techniques based on Table \ref{tab:main-results} 
may be useful to experimentalists wishing to observe specific patterns in the laboratory.

The results in this paper tie together many of the ideas explored in 
\cite{ss1999,sts2000,ts2002,pts2004} and provide an exhaustive description 
of the important resonant triad interactions for Faraday waves (with 
sufficiently weak damping).  Recent experiments used multi-frequency 
forcing of Faraday waves in order to control the transition between different 
nonlinear states and to suppress spatiotemporal disorder~\cite{ef2004}.  In 
particular, the authors of~\cite{ef2004} apply a perturbing third frequency to 
two-frequency forced patterns near a codimension-two point and interpret their 
results in terms of the temporal parities of the dominant forcing frequency and 
the perturbing frequency. Our results in Table \ref{tab:main-results} suggest 
that the frequencies themselves (not just the parity) and the forcing phases 
are important, thus providing an alternative approach for controlling patterns.

It will be interesting to extend our work to other systems. For example, in 
vertically vibrated convection, Boussinesq symmetry prohibits 
three-wave interactions \cite{rsbs2000}. Four-wave interactions are the 
important nonlinear interactions, and are the building blocks of complex 
square superlattice patterns observed in \cite{rsbp2000,rps2003}.   Applying 
techniques similar to those developed here might yield insight into this pattern 
selection mechanism as well.

\begin{acknowledgments}
We are grateful to Anne Catll\'{a} for a thorough reading of this manuscript 
and numerous helpful comments.  We thank Jude Higdon for assistance 
with the graphical preparation of several figures.  The work of CMT was 
supported by NSF grants DMS-9983726 and DMS-9983320.  The work 
of JP was supported by EPSRC grant GR/R52879/01.   The work of MS 
was supported by NASA grant NAG3-2364, NSF grant DMS-0309667, 
and by the NSF MRSEC Program under DMR-0213745. 
\end{acknowledgments}

\bibliography{tps2004}% Produces the bibliography via BibTeX.

\begin{thebibliography}{46}
\expandafter\ifx\csname natexlab\endcsname\relax\def\natexlab#1{#1}\fi
\expandafter\ifx\csname bibnamefont\endcsname\relax
  \def\bibnamefont#1{#1}\fi
\expandafter\ifx\csname bibfnamefont\endcsname\relax
  \def\bibfnamefont#1{#1}\fi
\expandafter\ifx\csname citenamefont\endcsname\relax
  \def\citenamefont#1{#1}\fi
\expandafter\ifx\csname url\endcsname\relax
  \def\url#1{\texttt{#1}}\fi
\expandafter\ifx\csname urlprefix\endcsname\relax\def\urlprefix{URL }\fi
\providecommand{\bibinfo}[2]{#2}
\providecommand{\eprint}[2][]{\url{#2}}

\bibitem[{\citenamefont{Faraday}(1831)}]{f1831}
\bibinfo{author}{\bibfnamefont{M.}~\bibnamefont{Faraday}},
  \bibinfo{journal}{Phil. Trans. R. Soc. Lond.} \textbf{\bibinfo{volume}{121}},
  \bibinfo{pages}{319} (\bibinfo{year}{1831}).

\bibitem[{\citenamefont{Miles and Henderson}(1990)}]{mh1990}
\bibinfo{author}{\bibfnamefont{J.}~\bibnamefont{Miles}} \bibnamefont{and}
  \bibinfo{author}{\bibfnamefont{D.}~\bibnamefont{Henderson}},
  \bibinfo{journal}{Ann. Rev. Fluid Mech.} \textbf{\bibinfo{volume}{22}},
  \bibinfo{pages}{143} (\bibinfo{year}{1990}).

\bibitem[{\citenamefont{\oneletter{H.W.} M{\"{u}}ller
  et~al.}(1998)\citenamefont{\oneletter{H.W.} M{\"{u}}ller, Friedrich, and
  Papathanassiou}}]{mfp1998}
\bibinfo{author}{\bibnamefont{\oneletter{H.W.} M{\"{u}}ller}},
  \bibinfo{author}{\bibfnamefont{R.}~\bibnamefont{Friedrich}},
  \bibnamefont{and}
  \bibinfo{author}{\bibfnamefont{D.}~\bibnamefont{Papathanassiou}}, in
  \emph{\bibinfo{booktitle}{Evolution of Spontaneous Structures in Dissipative
  Continuous Systems}}, edited by
  \bibinfo{editor}{\bibfnamefont{F.}~\bibnamefont{Busse}} \bibnamefont{and}
  \bibinfo{editor}{\bibnamefont{\oneletter{S.C.} M{\"{u}}ller}}
  (\bibinfo{publisher}{Springer}, \bibinfo{year}{1998}), Lecture Notes in
  Physics, pp. \bibinfo{pages}{231--265}.

\bibitem[{\citenamefont{\oneletter{H.W.} M{\"{u}}ller}(1993)}]{m1993}
\bibinfo{author}{\bibnamefont{\oneletter{H.W.} M{\"{u}}ller}},
  \bibinfo{journal}{Phys. Rev. Lett.} \textbf{\bibinfo{volume}{71}},
  \bibinfo{pages}{3287} (\bibinfo{year}{1993}).

\bibitem[{\citenamefont{\oneletter{W.S.} Edwards and Fauve}(1993)}]{ef1993}
\bibinfo{author}{\bibnamefont{\oneletter{W.S.} Edwards}} \bibnamefont{and}
  \bibinfo{author}{\bibfnamefont{S.}~\bibnamefont{Fauve}},
  \bibinfo{journal}{Phys. Rev. E} \textbf{\bibinfo{volume}{47}},
  \bibinfo{pages}{R788} (\bibinfo{year}{1993}).

\bibitem[{\citenamefont{\oneletter{W.S.} Edwards and Fauve}(1994)}]{ef1994}
\bibinfo{author}{\bibnamefont{\oneletter{W.S.} Edwards}} \bibnamefont{and}
  \bibinfo{author}{\bibfnamefont{S.}~\bibnamefont{Fauve}}, \bibinfo{journal}{J.
  Fluid Mech.} \textbf{\bibinfo{volume}{278}}, \bibinfo{pages}{123}
  (\bibinfo{year}{1994}).

\bibitem[{\citenamefont{Kudrolli et~al.}(1998)\citenamefont{Kudrolli, Pier, and
  \oneletter{J.P.} Gollub}}]{kpg1998}
\bibinfo{author}{\bibfnamefont{A.}~\bibnamefont{Kudrolli}},
  \bibinfo{author}{\bibfnamefont{B.}~\bibnamefont{Pier}}, \bibnamefont{and}
  \bibinfo{author}{\bibnamefont{\oneletter{J.P.} Gollub}},
  \bibinfo{journal}{Physica D} \textbf{\bibinfo{volume}{123}},
  \bibinfo{pages}{99} (\bibinfo{year}{1998}).

\bibitem[{\citenamefont{Arbell and Fineberg}(1998)}]{af1998}
\bibinfo{author}{\bibfnamefont{H.}~\bibnamefont{Arbell}} \bibnamefont{and}
  \bibinfo{author}{\bibfnamefont{J.}~\bibnamefont{Fineberg}},
  \bibinfo{journal}{Phys. Rev. Lett.} \textbf{\bibinfo{volume}{81}},
  \bibinfo{pages}{4384} (\bibinfo{year}{1998}).

\bibitem[{\citenamefont{Arbell and Fineberg}(2000{\natexlab{a}})}]{af2000a}
\bibinfo{author}{\bibfnamefont{H.}~\bibnamefont{Arbell}} \bibnamefont{and}
  \bibinfo{author}{\bibfnamefont{J.}~\bibnamefont{Fineberg}},
  \bibinfo{journal}{Phys. Rev. Lett.} \textbf{\bibinfo{volume}{84}},
  \bibinfo{pages}{654} (\bibinfo{year}{2000}{\natexlab{a}}).

\bibitem[{\citenamefont{Arbell and Fineberg}(2000{\natexlab{b}})}]{af2000b}
\bibinfo{author}{\bibfnamefont{H.}~\bibnamefont{Arbell}} \bibnamefont{and}
  \bibinfo{author}{\bibfnamefont{J.}~\bibnamefont{Fineberg}},
  \bibinfo{journal}{Phys. Rev. Lett.} \textbf{\bibinfo{volume}{85}},
  \bibinfo{pages}{756} (\bibinfo{year}{2000}{\natexlab{b}}).

\bibitem[{\citenamefont{Arbell and Fineberg}(2002)}]{af2002}
\bibinfo{author}{\bibfnamefont{H.}~\bibnamefont{Arbell}} \bibnamefont{and}
  \bibinfo{author}{\bibfnamefont{J.}~\bibnamefont{Fineberg}},
  \bibinfo{journal}{Phys. Rev. E} \textbf{\bibinfo{volume}{65}},
  \bibinfo{pages}{036244.1} (\bibinfo{year}{2002}).

\bibitem[{\citenamefont{Porter and Silber}(2002)}]{ps2002}
\bibinfo{author}{\bibfnamefont{J.}~\bibnamefont{Porter}} \bibnamefont{and}
  \bibinfo{author}{\bibfnamefont{M.}~\bibnamefont{Silber}},
  \bibinfo{journal}{Phys. Rev. Lett.} \textbf{\bibinfo{volume}{89}},
  \bibinfo{pages}{084501.1} (\bibinfo{year}{2002}).

\bibitem[{\citenamefont{Besson et~al.}(1996)\citenamefont{Besson,
  \oneletter{W.S.} Edwards, and Tuckerman}}]{bet1996}
\bibinfo{author}{\bibfnamefont{T.}~\bibnamefont{Besson}},
  \bibinfo{author}{\bibnamefont{\oneletter{W.S.} Edwards}}, \bibnamefont{and}
  \bibinfo{author}{\bibfnamefont{L.}~\bibnamefont{Tuckerman}},
  \bibinfo{journal}{Phys. Rev. E} \textbf{\bibinfo{volume}{54}},
  \bibinfo{pages}{507} (\bibinfo{year}{1996}).

\bibitem[{\citenamefont{Zhang and Vi{\~{n}}als}(1997{\natexlab{a}})}]{zv1997b}
\bibinfo{author}{\bibfnamefont{W.}~\bibnamefont{Zhang}} \bibnamefont{and}
  \bibinfo{author}{\bibfnamefont{J.}~\bibnamefont{Vi{\~{n}}als}},
  \bibinfo{journal}{J. Fluid Mech.} \textbf{\bibinfo{volume}{341}},
  \bibinfo{pages}{225} (\bibinfo{year}{1997}{\natexlab{a}}).

\bibitem[{\citenamefont{Lifshitz and \oneletter{D.M.} Petrich}(1997)}]{lp1997}
\bibinfo{author}{\bibfnamefont{R.}~\bibnamefont{Lifshitz}} \bibnamefont{and}
  \bibinfo{author}{\bibnamefont{\oneletter{D.M.} Petrich}},
  \bibinfo{journal}{Phys. Rev. Lett.} \textbf{\bibinfo{volume}{79}},
  \bibinfo{pages}{1261} (\bibinfo{year}{1997}).

\bibitem[{\citenamefont{Silber and \oneletter{M.R.E.} Proctor}(1998)}]{sp1998}
\bibinfo{author}{\bibfnamefont{M.}~\bibnamefont{Silber}} \bibnamefont{and}
  \bibinfo{author}{\bibnamefont{\oneletter{M.R.E.} Proctor}},
  \bibinfo{journal}{Phys. Rev. Lett} \textbf{\bibinfo{volume}{81}},
  \bibinfo{pages}{2450} (\bibinfo{year}{1998}).

\bibitem[{\citenamefont{Silber et~al.}(2000)\citenamefont{Silber,
  \oneletter{C.M.} Topaz, and \oneletter{A.C.} Skeldon}}]{sts2000}
\bibinfo{author}{\bibfnamefont{M.}~\bibnamefont{Silber}},
  \bibinfo{author}{\bibnamefont{\oneletter{C.M.} Topaz}}, \bibnamefont{and}
  \bibinfo{author}{\bibnamefont{\oneletter{A.C.} Skeldon}},
  \bibinfo{journal}{Physica D} \textbf{\bibinfo{volume}{143}},
  \bibinfo{pages}{205} (\bibinfo{year}{2000}).

\bibitem[{\citenamefont{\oneletter{D.P.} Tse
  et~al.}(2000)\citenamefont{\oneletter{D.P.} Tse, \oneletter{A.M.} Rucklidge,
  \oneletter{R.B.} Hoyle, and Silber}}]{trhs2000}
\bibinfo{author}{\bibnamefont{\oneletter{D.P.} Tse}},
  \bibinfo{author}{\bibnamefont{\oneletter{A.M.} Rucklidge}},
  \bibinfo{author}{\bibnamefont{\oneletter{R.B.} Hoyle}}, \bibnamefont{and}
  \bibinfo{author}{\bibfnamefont{M.}~\bibnamefont{Silber}},
  \bibinfo{journal}{Physica D} \textbf{\bibinfo{volume}{146}},
  \bibinfo{pages}{367} (\bibinfo{year}{2000}).

\bibitem[{\citenamefont{\oneletter{C.M.} Topaz and Silber}(2002)}]{ts2002}
\bibinfo{author}{\bibnamefont{\oneletter{C.M.} Topaz}} \bibnamefont{and}
  \bibinfo{author}{\bibfnamefont{M.}~\bibnamefont{Silber}},
  \bibinfo{journal}{Physica D} \textbf{\bibinfo{volume}{172}},
  \bibinfo{pages}{1} (\bibinfo{year}{2002}).

\bibitem[{\citenamefont{Porter et~al.}(in press)\citenamefont{Porter,
  \oneletter{C.M.} Topaz, and Silber}}]{pts2004}
\bibinfo{author}{\bibfnamefont{J.}~\bibnamefont{Porter}},
  \bibinfo{author}{\bibnamefont{\oneletter{C.M.} Topaz}}, \bibnamefont{and}
  \bibinfo{author}{\bibfnamefont{M.}~\bibnamefont{Silber}},
  \bibinfo{journal}{Phys. Rev. Lett.}  (\bibinfo{year}{in press}).

\bibitem[{\citenamefont{Zhang and Vi{\~{n}}als}(1997{\natexlab{b}})}]{zv1997a}
\bibinfo{author}{\bibfnamefont{W.}~\bibnamefont{Zhang}} \bibnamefont{and}
  \bibinfo{author}{\bibfnamefont{J.}~\bibnamefont{Vi{\~{n}}als}},
  \bibinfo{journal}{J. Fluid Mech.} \textbf{\bibinfo{volume}{336}},
  \bibinfo{pages}{301} (\bibinfo{year}{1997}{\natexlab{b}}).

\bibitem[{\citenamefont{Catll\'{a} et~al.}(2004)\citenamefont{Catll\'{a},
  Porter, and Silber}}]{cps2004}
\bibinfo{author}{\bibfnamefont{A.}~\bibnamefont{Catll\'{a}}},
  \bibinfo{author}{\bibfnamefont{J.}~\bibnamefont{Porter}}, \bibnamefont{and}
  \bibinfo{author}{\bibfnamefont{M.}~\bibnamefont{Silber}}
  (\bibinfo{year}{2004}), \bibinfo{note}{in prep.}

\bibitem[{\citenamefont{Silber and \oneletter{A.C.} Skeldon}(1999)}]{ss1999}
\bibinfo{author}{\bibfnamefont{M.}~\bibnamefont{Silber}} \bibnamefont{and}
  \bibinfo{author}{\bibnamefont{\oneletter{A.C.} Skeldon}},
  \bibinfo{journal}{Phys. Rev. E} \textbf{\bibinfo{volume}{59}},
  \bibinfo{pages}{5446} (\bibinfo{year}{1999}).

\bibitem[{\citenamefont{Porter and Silber}(2004)}]{ps2003}
\bibinfo{author}{\bibfnamefont{J.}~\bibnamefont{Porter}} \bibnamefont{and}
  \bibinfo{author}{\bibfnamefont{M.}~\bibnamefont{Silber}},
  \bibinfo{journal}{Physica D} \textbf{\bibinfo{volume}{190}},
  \bibinfo{pages}{93} (\bibinfo{year}{2004}).

\bibitem[{\citenamefont{Golubitsky et~al.}(1988)\citenamefont{Golubitsky,
  Stewart, and \oneletter{D.G.} Schaeffer}}]{gss1988}
\bibinfo{author}{\bibfnamefont{M.}~\bibnamefont{Golubitsky}},
  \bibinfo{author}{\bibfnamefont{I.}~\bibnamefont{Stewart}}, \bibnamefont{and}
  \bibinfo{author}{\bibnamefont{\oneletter{D.G.} Schaeffer}},
  \emph{\bibinfo{title}{Singularities and Groups in Bifurcation Theory: Vol.
  II}}, no.~\bibinfo{number}{69} in \bibinfo{series}{Appl. Math. Sci. Ser.}
  (\bibinfo{publisher}{Springer--Verlag}, \bibinfo{address}{New York},
  \bibinfo{year}{1988}).

\bibitem[{\citenamefont{\oneletter{S.T.} Milner}(1991)}]{m1991}
\bibinfo{author}{\bibnamefont{\oneletter{S.T.} Milner}}, \bibinfo{journal}{J.
  Fluid Mech.} \textbf{\bibinfo{volume}{225}}, \bibinfo{pages}{81}
  (\bibinfo{year}{1991}).

\bibitem[{\citenamefont{\oneletter{V.E.} Zakharov}(1968)}]{z1968}
\bibinfo{author}{\bibnamefont{\oneletter{V.E.} Zakharov}}, \bibinfo{journal}{J.
  Appl. Mech. Tech. Phys.} \textbf{\bibinfo{volume}{2}}, \bibinfo{pages}{190}
  (\bibinfo{year}{1968}).

\bibitem[{\citenamefont{\oneletter{J.W} Miles}(1977)}]{m1977}
\bibinfo{author}{\bibnamefont{\oneletter{J.W} Miles}}, \bibinfo{journal}{J.
  Fluid Mech.} \textbf{\bibinfo{volume}{83}}, \bibinfo{pages}{153}
  (\bibinfo{year}{1977}).

\bibitem[{\citenamefont{\oneletter{L.J.F} Broer}(1974)}]{b1974}
\bibinfo{author}{\bibnamefont{\oneletter{L.J.F} Broer}},
  \bibinfo{journal}{Appl. Sci. Res.} \textbf{\bibinfo{volume}{29}},
  \bibinfo{pages}{430} (\bibinfo{year}{1974}).

\bibitem[{\citenamefont{Lyngshansen and Alstrom}(1997)}]{la1997}
\bibinfo{author}{\bibfnamefont{P.}~\bibnamefont{Lyngshansen}} \bibnamefont{and}
  \bibinfo{author}{\bibfnamefont{P.}~\bibnamefont{Alstrom}},
  \bibinfo{journal}{J. Fluid Mech.} \textbf{\bibinfo{volume}{351}},
  \bibinfo{pages}{301} (\bibinfo{year}{1997}).

\bibitem[{\citenamefont{\oneletter{A.C.} Radder}(1992)}]{r1992}
\bibinfo{author}{\bibnamefont{\oneletter{A.C.} Radder}}, \bibinfo{journal}{J.
  Fluid Mech.} \textbf{\bibinfo{volume}{237}}, \bibinfo{pages}{435}
  (\bibinfo{year}{1992}).

\bibitem[{\citenamefont{Krasitskii}(1994)}]{k1994}
\bibinfo{author}{\bibfnamefont{V.~P.} \bibnamefont{Krasitskii}},
  \bibinfo{journal}{J. Fluid Mech.} \textbf{\bibinfo{volume}{272}},
  \bibinfo{pages}{1} (\bibinfo{year}{1994}).

\bibitem[{\citenamefont{Mahr et~al.}(1996)\citenamefont{Mahr, Groisman, and
  Rehberg}}]{mgr1996}
\bibinfo{author}{\bibfnamefont{T.}~\bibnamefont{Mahr}},
  \bibinfo{author}{\bibfnamefont{A.}~\bibnamefont{Groisman}}, \bibnamefont{and}
  \bibinfo{author}{\bibfnamefont{I.}~\bibnamefont{Rehberg}},
  \bibinfo{journal}{J. Magn. Magn. Mater.} \textbf{\bibinfo{volume}{159}},
  \bibinfo{pages}{L45} (\bibinfo{year}{1996}).

\bibitem[{\citenamefont{Bechhoefer and Johnson}(1996)}]{bj1996}
\bibinfo{author}{\bibfnamefont{J.}~\bibnamefont{Bechhoefer}} \bibnamefont{and}
  \bibinfo{author}{\bibfnamefont{B.}~\bibnamefont{Johnson}},
  \bibinfo{journal}{Am. J. Phys.} \textbf{\bibinfo{volume}{64}},
  \bibinfo{pages}{1482} (\bibinfo{year}{1996}).

\bibitem[{\citenamefont{Kumar and \oneletter{L.S.} Tuckerman}(1994)}]{kt1994}
\bibinfo{author}{\bibfnamefont{K.}~\bibnamefont{Kumar}} \bibnamefont{and}
  \bibinfo{author}{\bibnamefont{\oneletter{L.S.} Tuckerman}},
  \bibinfo{journal}{J. Fluid Mech.} \textbf{\bibinfo{volume}{279}},
  \bibinfo{pages}{49} (\bibinfo{year}{1994}).

\bibitem[{\citenamefont{Dionne and Golubitsky}(1992)}]{dg1992}
\bibinfo{author}{\bibfnamefont{B.}~\bibnamefont{Dionne}} \bibnamefont{and}
  \bibinfo{author}{\bibfnamefont{M.}~\bibnamefont{Golubitsky}},
  \bibinfo{journal}{Z. Angew. Math. Phys.} \textbf{\bibinfo{volume}{43}},
  \bibinfo{pages}{36} (\bibinfo{year}{1992}).

\bibitem[{\citenamefont{Dionne et~al.}(1997)\citenamefont{Dionne, Silber, and
  \oneletter{A.C.} Skeldon}}]{dss1997}
\bibinfo{author}{\bibfnamefont{B.}~\bibnamefont{Dionne}},
  \bibinfo{author}{\bibfnamefont{M.}~\bibnamefont{Silber}}, \bibnamefont{and}
  \bibinfo{author}{\bibnamefont{\oneletter{A.C.} Skeldon}},
  \bibinfo{journal}{Nonlinearity} \textbf{\bibinfo{volume}{10}},
  \bibinfo{pages}{321} (\bibinfo{year}{1997}).

\bibitem[{\citenamefont{Christiansen et~al.}(1992)\citenamefont{Christiansen,
  Alstrom, and \oneletter{M.T.} Levinsen}}]{cal1992}
\bibinfo{author}{\bibfnamefont{B.}~\bibnamefont{Christiansen}},
  \bibinfo{author}{\bibfnamefont{P.}~\bibnamefont{Alstrom}}, \bibnamefont{and}
  \bibinfo{author}{\bibnamefont{\oneletter{M.T.} Levinsen}},
  \bibinfo{journal}{Phys. Rev. Lett.} \textbf{\bibinfo{volume}{68}},
  \bibinfo{pages}{2157} (\bibinfo{year}{1992}).

\bibitem[{\citenamefont{\oneletter{A.C.} Newell and Pomeau}(1993)}]{np1993}
\bibinfo{author}{\bibnamefont{\oneletter{A.C.} Newell}} \bibnamefont{and}
  \bibinfo{author}{\bibfnamefont{Y.}~\bibnamefont{Pomeau}},
  \bibinfo{journal}{J. Phys. A--Math Gen.} \textbf{\bibinfo{volume}{26}},
  \bibinfo{pages}{L429} (\bibinfo{year}{1993}).

\bibitem[{\citenamefont{Echebarria and Riecke}(2001)}]{er2001}
\bibinfo{author}{\bibfnamefont{B.}~\bibnamefont{Echebarria}} \bibnamefont{and}
  \bibinfo{author}{\bibfnamefont{H.}~\bibnamefont{Riecke}},
  \bibinfo{journal}{Physica D} \textbf{\bibinfo{volume}{158}},
  \bibinfo{pages}{45} (\bibinfo{year}{2001}).

\bibitem[{\citenamefont{\oneletter{A.M.} Rucklidge and \oneletter{W.J.}
  Rucklidge}(2003)}]{rr2003}
\bibinfo{author}{\bibnamefont{\oneletter{A.M.} Rucklidge}} \bibnamefont{and}
  \bibinfo{author}{\bibnamefont{\oneletter{W.J.} Rucklidge}},
  \bibinfo{journal}{Physica D} \textbf{\bibinfo{volume}{78}},
  \bibinfo{pages}{62} (\bibinfo{year}{2003}).

\bibitem[{\citenamefont{Chossat}(2001)}]{c2001}
\bibinfo{author}{\bibfnamefont{P.}~\bibnamefont{Chossat}},
  \bibinfo{journal}{Dynam. Cont. Dis. Ser. A} \textbf{\bibinfo{volume}{8}},
  \bibinfo{pages}{575} (\bibinfo{year}{2001}).

\bibitem[{\citenamefont{Epstein and Fineberg}(in press)}]{ef2004}
\bibinfo{author}{\bibfnamefont{T.}~\bibnamefont{Epstein}} \bibnamefont{and}
  \bibinfo{author}{\bibfnamefont{J.}~\bibnamefont{Fineberg}},
  \bibinfo{journal}{Phys. Rev. Lett.}  (\bibinfo{year}{in press}).

\bibitem[{\citenamefont{\oneletter{J.L.} Rogers
  et~al.}(2000{\natexlab{a}})\citenamefont{\oneletter{J.L.} Rogers,
  \oneletter{M.F.} Schatz, \oneletter{J.L.} Bougie, and \oneletter{J.B.}
  Swift}}]{rsbs2000}
\bibinfo{author}{\bibnamefont{\oneletter{J.L.} Rogers}},
  \bibinfo{author}{\bibnamefont{\oneletter{M.F.} Schatz}},
  \bibinfo{author}{\bibnamefont{\oneletter{J.L.} Bougie}}, \bibnamefont{and}
  \bibinfo{author}{\bibnamefont{\oneletter{J.B.} Swift}},
  \bibinfo{journal}{Phys. Rev. Lett.} \textbf{\bibinfo{volume}{84}},
  \bibinfo{pages}{87} (\bibinfo{year}{2000}{\natexlab{a}}).

\bibitem[{\citenamefont{\oneletter{J.L.} Rogers
  et~al.}(2000{\natexlab{b}})\citenamefont{\oneletter{J.L.} Rogers,
  \oneletter{M.F.} Schatz, Brausch, and Pesch}}]{rsbp2000}
\bibinfo{author}{\bibnamefont{\oneletter{J.L.} Rogers}},
  \bibinfo{author}{\bibnamefont{\oneletter{M.F.} Schatz}},
  \bibinfo{author}{\bibfnamefont{O.}~\bibnamefont{Brausch}}, \bibnamefont{and}
  \bibinfo{author}{\bibfnamefont{W.}~\bibnamefont{Pesch}},
  \bibinfo{journal}{Phys. Rev. Lett.} \textbf{\bibinfo{volume}{85}},
  \bibinfo{pages}{4281} (\bibinfo{year}{2000}{\natexlab{b}}).

\bibitem[{\citenamefont{\oneletter{J.L.} Rogers
  et~al.}(2003)\citenamefont{\oneletter{J.L.} Rogers, Pesch, and
  \oneletter{M.F.} Schatz}}]{rps2003}
\bibinfo{author}{\bibnamefont{\oneletter{J.L.} Rogers}},
  \bibinfo{author}{\bibfnamefont{W.}~\bibnamefont{Pesch}}, \bibnamefont{and}
  \bibinfo{author}{\bibnamefont{\oneletter{M.F.} Schatz}},
  \bibinfo{journal}{Nonlinearity} \textbf{\bibinfo{volume}{16}},
  \bibinfo{pages}{C1} (\bibinfo{year}{2003}).

\end{thebibliography}

\end{document}